\documentclass[11pt,a4paper]{article}
\usepackage{jheppub,bm,color}
\usepackage{amsmath,amssymb,bm}

\newcommand{\na}{{\bm \nabla}}
\newcommand{\dd}{{\rm d}}
\newcommand{\xv}{{\bm x}}

\newcommand{\Bv}{{\bm B}}
\newcommand{\Ev}{{\bm E}}

\newcommand{\kv}{{\bm k}}

\newcommand{\qv}{{\bm q}}

\newcommand{\vv}{{\bm v}}
\newcommand{\zev}{{\bm 0}}

\newcommand{\rmA}{{\rm A}}
\newcommand{\rmV}{{\rm V}}

\newcommand{\rma}{{\rm a}}
\newcommand{\rmr}{{\rm r}}

\newcommand{\e}{{\rm e}}
\renewcommand{\i}{{\rm i}}

\newcommand{\mL}{{\mathcal L}}

\def \pd {\partial}

\def \le {\left}
\def \ri {\right}

\def \vp {\bm{p}}
\def \vx {\bm{x}}
\def \vk {\bm{k}}
\def \vq {\bm{q}}

\def \vB {\bm{B}}

\def \G {\Gamma}
\def \eff {\rm eff}
\def \CG {{\cal G}}
\def \sL {{\cal L}}

\def \a {\alpha}
\def \b {\beta}

\def \d {\delta}
\def \g {\gamma}

\def \o {\omega}

\def \l {\lambda}

\def \s {\sigma}

\def \CC {{\cal C}}

\def \vB {\bm{B}}

\def \no {\nonumber}
\def \bes {\begin{subequations} }
\def \ees {\end{subequations}}

\def \<{\langle}
\def \>{\rangle}

\usepackage[normalem]{ulem}  

\begin{document}
\title{Positive magnetoresistance induced by hydrodynamic fluctuations in chiral media}
\author[a]{Noriyuki~Sogabe,}
\author[b]{Naoki Yamamoto}
\author[a]{and Yi Yin}

\affiliation[a]{Quark Matter Research Center, Institute of Modern Physics, Chinese Academy of Sciences, Lanzhou, Gansu, 073000, China }

\affiliation[b]{Department of Physics, Keio University, Yokohama 223-8522, Japan}

\abstract{We analyze the combined effects of hydrodynamic fluctuations and chiral magnetic effect (CME) for a chiral medium in the presence of a background magnetic field. Based on the recently developed non-equilibrium effective field theory, we show fluctuations give rise to a CME-related positive contribution to magnetoresistance, while the early studies without accounting for the fluctuations find a CME-related negative magnetoresistance. At zero axial relaxation rate, the fluctuations contribute to the transverse conductivity in addition to the longitudinal one.}

\emailAdd{nori.sogabe@gmail.com}
\emailAdd{nyama@rk.phys.keio.ac.jp}
\emailAdd{yiyin@impcas.ac.cn}

\maketitle

\section{Introduction}
The transport properties of a chiral medium (many-body system involving chiral fermions) and their deep connection to quantum anomalies have attracted significant interests recently. 
Of particular importance is the behavior of electric conductivity (or its inverse, electric resistance) under the external magnetic field $\Bv$. 
In table-top experiments, the negative magnetoresistance is proposed as a signature of the chiral magnetic effect (CME), the anomaly-induced vector current in the presence of magnetic field and chiral charge imbalance \cite{Fukushima:2008xe,Kharzeev:2007jp,Nielsen:1983rb,Vilenkin:1980fu}. 
Indeed, as shown in refs.~\cite{Son:2012bg,Stephanov:2014dma,Fukushima:2019ugr}, the balance between the axial charge density $n_{\rm A}$ production due to the chiral anomaly and axial charge relaxation requires that in a steady state in the presence of electric field $\Ev$, the axial charge density $n_{\rm A}\propto C {\Ev}\cdot {\Bv}/r$, where $r$ denotes the axial charge relaxation rate and $C$ is the anomaly coefficient. Then, with the CME, one finds an additional contribution to the longitudinal conductivity
\begin{align}
\label{CME-L}
\Delta \sigma_{\rm L}\propto \frac{C^{2}{\bm B}^2}{r}\, . 
\end{align}
The measurements of magnetoresistance in Weyl and Dirac semimetals have been reported in refs.~\cite{Li:2014bha,Xiong413,Huang:2015eia,Arnold_2016}. 

Nevertheless, the fluctuation effects have not yet been taken into account in eq.~\eqref{CME-L}. 
It is well-known that in an ordinary fluid, 
the fluctuations and interactions among sound and diffusive modes lead to significant effects on the behavior of transport coefficients \cite{POMEAU197563,Kovtun:2003vj,Kovtun:2012rj,PhysRevA.16.732,Kovtun:2014nsa,Grossi:2020ezz,Grossi:2021gqi}. 
Therefore, one may naturally ask how fluctuations would modify the magnetoresistance in a chiral medium. 
Addressing this question is the primary goal of the present work. 
For definiteness, we shall consider the fluctuations of both vector and axial charge densities.
As in previous studies~\cite{Son:2012bg,Hattori:2017usa}, we assume that $r$ is parametrically small compared with the microscopic relaxation rate, and hence we include the axial charge density as a slow mode. 

We here use the recently developed non-equilibrium effective field theory (EFT) for hydrodynamics fluctuations~\cite{Crossley:2015,Crossley:2017} (see ref.~\cite{Glorioso:2018wxw} for a review and refs.~\cite{Haehl:2015foa,Jensen:2017kzi} for related developments), including the effects of quantum anomaly~\cite{Glorioso:2017lcn}, to perform our analysis. 
Compared with the traditional methods, the EFTs are derived based on the symmetries and action principle and provide a basis for the systematic analysis. 
In some situations, such as the one considered in ref.~\cite{Chen-Lin:2018kfl}, EFT calculations lead to different results as compared with traditional analysis. 
Previous work including the fluctuations of a single chiral charge and CME can be found in ref.~\cite{Delacretaz:2020jis}.
See refs.~\cite{Fukushima:2017lvb,Fukushima:2019ugr} for the diagrammatic calculation of magnetoresistance for quark-gluon plasma (QGP) based on perturbative QCD. 

In two situations, $r \neq 0$ and $r=0$, we determine specific corrections to the conductivity due to the combined effects of the CME and fluctuations in the small $\Bv$ regime (see eq.~\eqref{eq:Sigma-r-fin} and eq.~\eqref{sigma-zeror}, respectively).
Physically, those CME-related contributions have two origins. 
First, the CME is proportional to the axial chemical potential which generically depends on charge density non-linearly and gives rise to non-linear coupling among density fluctuations. 
Second, the CME modifies the dispersion relation of fluctuations modes~\cite{Kharzeev:2010gd,Stephanov:2014dma}.
Given the difference in physical origin, 
we should not be surprised to see that
the fluctuation corrections are in marked difference from eq.~\eqref{CME-L}. 
One important qualitative feature we observe is that the sign of fluctuation contributions is opposite to that of eq.~\eqref{CME-L}, meaning they give rise to positive magnetoresistance. 
Moreover, we find a non-zero contribution to transverse conductivity when $r=0$. 
As already noticed in some references~\cite{Baumgartner:2017kme,Fukushima:2017lvb,Fukushima:2019ugr}, 
other mechanisms unrelated to the anomaly could cause magnetoresistance. 
The present work aims to demonstrate that even if one only focuses on the effects of the chiral anomaly, the contribution from fluctuations to magnetoresistance can be qualitatively different from that at ``tree-level.''
Our results might apply to physical systems, such as the QGP created by heavy-ion collisions, Weyl semimetals, and the electroweak plasma in the primordial Universe.

This paper is organized as follows.  
After reviewing the construction of the EFT action in section~\ref{sec:EFT}, 
we determine the relevant Feynman rules and vertices. 
In sections~\ref{sec:con} and \ref{sec:zeror}, we respectively calculate the conductivity at one-loop at finite and vanishing axial charge relaxation rate. 
We conclude in section~\ref{sec:conclusion}. 

In this paper, we use $\hbar=c=k_{\rm B}=e=1$ and the mostly plus metric $\eta^{\mu\nu} ={\rm diag}\, (-1,1,1,1)$.
We use the shorthand notation for space-time and frequency-momentum integrations:$\int_x = \int {\rm d}^4 x$ with $x^\mu=(t,\xv)$; $\int_Q=\int_{q^0}\int_\qv$ with $Q^\mu=(q^0,\qv)$ and $q\equiv|\qv|$; $\int_{q^0}=\int \dd q^0 /(2\pi)$; $\int_{\vq}=\int \dd ^3\vq/(2\pi)^{3}$.

\section{Non-equilibrium effective field theory}
\label{sec:EFT}

\subsection{The action}
\label{sec:the action}
We are interested in the fluctuation dynamics of vector charge density  $n_{\rm V}$ and axial charge density $n_{\rm A}$ in a chiral medium. As already mentioned in the Introduction,
we shall assume the relaxation rate of $n_{\rm A}$ is small compared with the microscopic relaxation rate. Furthermore, we shall limit ourselves to situations that temperature is much smaller than vector chemical potential $\mu_{\rm V}$ and/or axial chemical potential $\mu_{\rm A}$. 
In this regime, we could ignore the mixing of $n_{\rm V}$ and $n_{\rm A}$ with the energy density. We also note in electron systems including Weyl semimetals, the mean free path of momentum-relaxing scattering (e.g., impurity scattering) can typically be shorter than the mean free path of momentum-conserving scattering (electron-electron scattering). In such a situation, the momentum is not a hydrodynamic variable, and ignoring the coupling of (charge) density modes to sound/shear modes can be well justified. Therefore, in long-time and large-distance limits, we can integrate out other modes and obtain the effective action $I_{{\rm eff}}$ describing the remaining slow modes $n_{\rm V}$ and $n_{\rm A}$.
In general, it is difficult to obtain $I_{\eff}$ directly from microscopic theories. 
Instead, one should construct $I_{\eff}$ based on the symmetries together with other physical requirements, as we shall do below following the formalism developed by refs.~\cite{Crossley:2015,Crossley:2017,Glorioso:2017lcn} (see ref.~\cite{Glorioso:2018wxw} for a pedagogical introduction). 

We begin with the path integral representation of the generating functional on the Schwinger-Keldysh contour,
\begin{align}
\label{I-Z}
\e^{{\rm i} W[A^\rmr,A^\rma]} = 
\int  \prod_{\alpha={\rm V,A}}\prod_{s=\rma,\rmr} \,{\cal D} \psi_{\a}^s \,  
\e^{\i I_{{\rm eff}}[\psi^\rma,\psi^\rmr;A^{\rm a},A^{\rm r}] } \, .
\end{align}
Here, we have introduced the external gauge fields $A^{\rm r}_\alpha$ and $A^{\rm a}_\alpha$ and the dynamical fields $\psi^{\rm r}_\alpha$ and $\psi^{\rm a}_\alpha$ associated with charge density $n_{\alpha}$, in the ``r-'' and ``a-'' basis. 
One can interpret $\psi^{\rmr}_\alpha$ and $ \psi^{\rma}_\alpha$ as the $\rm{U}(1)$ phase rotations of each fluid element (see refs.~\cite{Crossley:2015,Glorioso:2018wxw} for more details). 
The $\rm r$-variables are related to the physical observables, while the $\rm a$-variables are the associated noise variables. 

Next, we list various symmetries and consistency requirements which $I_{{\rm eff}}$ should satisfy:
\begin{enumerate}

\item
Gauge symmetries: $I_{\eff}$ has to be invariant under ${\rm U}(1)_{\rm V}$ gauge symmetry. Furthermore, we require ${\rm U}(1)_{\rm A}$ gauge symmetry for $I_{\eff}$ in the limit that the axial relaxation and the chiral anomaly are absent. 
Here, 
${\rm U}(1)_{\alpha}$ gauge transformation can be written explicitly as
\begin{align}
\label{gauge}
A^{s}_{\mu,\alpha} \to A^{s}_{\mu,\alpha} - \pd_{\mu}\phi_\alpha^s\, , 
\quad
\psi_\alpha^s \to \psi_\alpha^s +\phi_\alpha^s \, ,
\end{align}
where $\phi_{\a}^s$ is an arbitrary ${\rm U}(1)_{\a}$ phase. For a term invariant under the ${\rm U}(1)_{\a}$ symmetry, its dependence on $A^{\mu,s}_\alpha$ and $\psi_\alpha^s$ should come through the gauge-invariant combination: 
\begin{align}
\label{C-def}
{\cal A}^{s}_{\mu,\alpha}= A^{s}_{\mu,\alpha} +\pd_{\mu}\psi_\alpha^s\, .
\end{align}
In particular, the vector and axial chemical potentials are expressed as 
\begin{align}
\label{mu-def}
\mu_{\alpha} = {\cal A}_{0, \alpha}^{\rm r}.
\end{align}

\item Shift symmetries: 
For each fluid element, it should have the freedom of making independent ${\rm U}(1)_{\rm \a}$ phase rotations as far as those phases $ \zeta_{\a}(\vx)$ are time-independent: 
\begin{align}
\label{shift-sym}
\psi_\alpha^\rmr \rightarrow \psi_\alpha^\rmr + \zeta _{\alpha}(\xv) \, . 
\end{align}
Note that shift symmetries will be absent when the global ${\rm U}(1)_{\rm \a}$ symmetry is spontaneously broken (see refs.~\cite{Dubovsky:2011sj,Crossley:2015} for further details). 
\item Dynamical Kubo-Martin-Schwinger (KMS) symmetry: 
Suppose the microscopic theory is invariant under a $\mathbb{ Z}_{2}$  anti-unitary transformation $\Theta$,
then $I_{{\rm eff}}$ is invariant under the KMS transformation \cite{Crossley:2017}, which, in the ``classical'' limit that quantum fluctuations are small compared with the thermodynamic fluctuations, is defined as
\begin{subequations}
\begin{align}
\label{KMS-r}
{\cal A}^{\rmr}_\mu & \rightarrow \Theta {\cal A}^{\rmr}_\mu\,,
\quad
{\cal A}^{\rma}_{\mu} \rightarrow \Theta {\cal A}^{\rma}_\mu+ \frac{\rm i}{T} \Theta \pd_{t}{\cal A}^{\rmr}_\mu\, ,
\\
\label{KMS-a}
\mu_{\a}&\rightarrow \Theta \mu_{\alpha}\,, 
\quad
\psi_{\alpha}^\rma\rightarrow\Theta \psi_{\a}^\rma+ \frac{\rm i}{T} \Theta\pd_{t}\psi_{\a}^\rmr \, ,
\end{align}
\end{subequations}
where $T$ is the background temperature. 
The dynamical KMS symmetry is motivated by the KMS condition satisfied by a thermal system and can be viewed as a definition of local thermal equilibrium. Generically, one can take $\Theta$ that includes $\cal T$, i.e., $\Theta$ can be $\cal T$ itself, or any combination of ${\cal C},{\cal P}$ with ${\cal T}$~\cite{Crossley:2017}, depending on the systems of interest. 
The presence of background magnetic field and vector charge will break the symmetries under ${\mathcal T}$ and ${\mathcal C}$, respectively, so we shall take $\Theta = \mathcal C \mathcal P \mathcal T$ in this work. 

\item Unitarity: 
The unitarity of the underlying system requires that (suppressing $\a$ and $\mu$ indices)
\begin{align}
\label{unitarity}
&\, I_{\rm eff}[\psi^\rmr,A^\rmr;\psi^\rma=0,A^\rma=0]=0\,, \\
&\, I^{*}_{{\rm eff}}[\psi^\rmr,A^\rmr;\psi^\rma,A^\rma]=-I_{\rm eff}[\psi^\rmr,A^\rmr;-\psi^\rma,-A^\rma]\, . 
\end{align}

\end{enumerate}

To consider the low-energy regime of the system, we also perform a derivative expansion based on the basic philosophy of EFT. For definiteness, we take the following counting scheme in this paper: ${\cal A}_{0,\alpha}^{\rm a} \sim \pd_t {\cal A}_{0,\alpha}^{\rm r} \sim \epsilon^2$, ${\cal A}_{i,\alpha}^{\rm a} \sim \pd_t {\cal A}_{i,\alpha}^{\rm r} \sim \epsilon$ (such that $\pd_t \psi_{\alpha}^{\rm a} \sim \epsilon^2$, ${\bm \nabla} \psi_{\alpha}^{\rm a} \sim \epsilon$), and $\psi_{\alpha}^{\rm a} \sim \pd_t \psi_{\alpha}^{\rm r} \sim \epsilon^0$, where $\epsilon$ is a small expansion parameter.

Now, we are ready to write down the non-equilibrium action $I_{{\rm eff}}$ explicitly. 
Because of eq.~\eqref{unitarity}, $\sL$ has to contain at least one power of $\rm a$-field. 
We shall study $\sL$ up to quadratic order in $\rm a$-field.
More explicitly, we consider the Lagrangian density $\sL$, which is related to $I_{\rm eff}$ through the standard relation
\begin{align}
I_{\rm eff}[\psi^\rmr,A^\rmr;\psi^\rma,A^\rma]=\int_x {\sL}[\psi^\rmr,A^\rmr;\psi^\rma,A^\rma]\, ,
\end{align}
and divide $\sL$ into three parts:
\begin{align}
\label{L-pieces}
\sL = \sL_{\rm inv}+\sL_{\rm damp}+\sL_{\rm anom}\,.
\end{align}
Here, $\sL_{\rm inv}$ corresponds to $\sL$ in the limit that both the effects of the axial charge damping and chiral anomaly are absent. 
In this case, $n_{\rm A}$ is also conserved. Hence, $\mathcal L_{\rm inv}$ up to ${\cal O}(\epsilon^2)$ should be of the same form as the hydrodynamic effective action with two conserved charges as derived in ref.~\cite{Crossley:2015} (see also ref.~\cite{Chen-Lin:2018kfl}):
\begin{align}
\label{L-norm-V}
\mL_{\rm inv}&= \sum_\alpha n_\alpha {\cal A}_{0,\alpha}^{\rma} - \sum_{i,j,\alpha,\b} \sigma^{ij}_{\alpha\b} 
\left( {\cal A}^{\rm a}_{i,\alpha}  \pd_{t}{\cal A}^{\rm r}_{j,\beta}  
- \i T {\cal A}^\rma_{i,\alpha} {\cal A}^\rma_{j,\beta}  \right) \, ,
\end{align}
where $\s^{ij}_{\a\b} $ is the conductivity matrix which is symmetric with respect to $(i,j)$ and $(\alpha,\beta)$. 
Because of the shift symmetries, 
$\sL_{\rm inv}$ is independent of $\mathcal {A}^{\rmr}_{i,\alpha}$. 

Turning to $\sL_{\rm anom}$, which describes the effects of the chiral anomaly, we explicitly have
\begin{align}
\label{L-anom}
\mL_{\rm anom}& = C \psi_\rmA^\rma \Ev\cdot\Bv + C \mu_{\rmV}  \Bv \cdot \na \psi_\rmA^\rma + C \mu_{\rmA} \Bv \cdot \bm{ {\cal A}}_{\rmV}^\rma \,, 
\end{align}
where $C=1/(2\pi^2)$ denotes the anomaly coefficient and the electric and magnetic fields are defined by $\Ev = \na A^\rmr_{0,\rmV} - \partial_t \bm{A} ^\rmr_\rmV$ and $\Bv = \na \times \bm{A}^\rmr_\rmV$. To simplify the expression, we shall consider the cases in the absence of the axial gauge fields here and from now on. The first term in eq.~\eqref{L-anom} leads to the anomaly contribution to the non-conservation of the axial current (see eq.~(\ref{dJA}) below). 
The second and the third terms correspond to the chiral separation effect (CSE) \cite{Son:2004tq,Metlitski:2005pr} and CME, respectively. 
In appendix~\ref{sec:Anomaly-matching}, we present the derivation of eq.~\eqref{L-anom} by generalizing the formulation for a single chiral charge in ref.~\cite{Glorioso:2017lcn}.

Finally, we {\it postulate} to use
\begin{align}
\label{L-norm-A}
\mL_{\rm damp}&= 
-\g \left[ \mu_{\rmA} \psi_\rmA^\rma - \i T (\psi_\rmA^\rma )^2 \right] \,,
\end{align}
to describe the axial charge relaxation. Here, $\g$ denotes the axial damping coefficient, which is assumed to be ${\cal O}(\epsilon^2)$ so that $\mathcal L_{\rm damp}$ contributes to the same order as the other terms in eq.~(\ref{L-pieces}). 
Equation~\eqref{L-norm-A} satisfies all requirements as listed above. 

We here point out that the equations of motion from $I_{\rm eff}$ is equivalent to the (non-)conservation equations for the currents. 
For the vector charge density, we have
\begin{align}
\label{eq:con-noncon}
\frac{\delta I_{\rm eff}}{\delta \psi^\rma_{\rm V}} =0
\leftrightarrow\pd_{\mu}J^{\mu, \rmr}=0\, ,
\quad
J^{\mu,\rmr} = \frac{\delta I_{\rm eff} }{\delta A^{\rma}_{\mu,{\rm V}}}\,,
\end{align}
since $I_{\rm eff}$ only depends on the combination $A^{\rm a}_{\mu,{\rm V}}+\pd_{\mu}\psi^{\rm a}_{\rm V}$ but not on $A^{\rm a}_{\mu,{\rm V}}$ and $\psi^{\rm a}_{\rm V}$ individually.  
For later purpose, we obtain the expressions for $\bm{J}^\rmr$ and $\bm{J}^\rma$ by differentiating (\ref{L-pieces}) with respect to $\bm{A}^\rma$ and $\bm{A}^\rmr$, respectively:
\begin{align}
\label{jr}
{\bm J}^{\rm r} &= - \s_{\rm V} \na \mu_{\rm V}+ 2 \i T \s_{\rm V}\na \psi_{\rm V}+C\mu_{\rm A}\vB\,, \\
\label{ja}
{\bm J}^{\rm a} &=\pd_{t}\le(\s_{\rm V}\na \psi_{\rm V}+C\psi_{\rm A} \vB \ri)+ C\na \times \left(\mu_{\rm A}\na \psi_{\rmV}+\mu_{\rmV}\na \psi_{\rm A}+{\bm E} \psi_A \right)\,.
\end{align}
We can also define the axial current $J^{\mu}_{\rm A}$ through the variation of $I_{\rm eff}$. In that case, we find that the equation of motion for $\psi^{\rm a}_{\rm A}$ is nothing but the non-conservation equation for the axial current due to the chiral anomaly and axial charge damping,
\begin{align}
\label{dJA}
\pd_{\mu}J^{\mu,\rmr}_{\rm A}=C {\Ev}\cdot {\Bv}- \g \mu_{\rm A}+2 {\rm i} \g T \psi^{\rma}_{\rm A} \,  . 
\end{align}

In summary, we shall use the following effective action up to ${\cal O}(\epsilon^2)$ for the subsequent analysis for systems with a background magnetic field $\Bv$ based on eqs.~\eqref{L-norm-V}, \eqref{L-anom}, and \eqref{L-norm-A}:
\begin{align}
\label{action}
\sL &= \sum_{\alpha}
(\pd_{t}\psi_{\alpha}) n_{\alpha}
-\sum_{\alpha}\sigma_\alpha \le[ ({\bm \nabla}\psi_\alpha) \cdot {\bm \nabla}\mu_\alpha - {\rm i} T ({\bm \nabla}\psi_\alpha)^2\ri] \no \\
&\quad 
+C \mu_{\rm A}\vB\cdot{\bm \nabla} \psi_{\rm V}+C \mu_{\rm V}\vB\cdot {\bm \nabla} \psi_{\rm A}
-\g \le[ \psi_{\rm A} \mu_{\rm A}-{\rm i} T (\psi_{\rm A})^{2}\ri]\,.
\end{align}
For notational brevity, here and hereafter, we suppress the ${\rm a}$-index for $\psi^\rma_{\a}$. Note that ${\bm B} = {\cal O}(\epsilon)$ in our counting scheme above, and we may ignore the ${\bm B}$-dependence of $\sigma_{\alpha}$ and $\gamma$ at the level of this effective Lagrangian.

In this work, our goal is to showcase the non-trivial interplay among the axial charge density relaxation, CME, and fluctuations in the simplest possible settings. 
In eq.~\eqref{action}, we have assumed that at the tree level, $\s^{ij}_{\a\b}=\s_{\a}\,\d^{ij}\d_{\a\b}$, which is sufficient for the present illustrative purpose. 
In the same spirit, we shall use the susceptibility matrix $\chi_{\a\b}\equiv \pd n_{\a}/\pd \mu_{\b}$ which is also diagonalized, $\chi_{\a\b}=\chi_{\a}\delta_{\a\b}$.

Before closing this section, we point out that the first equation in eq.~(\ref{eq:con-noncon}) and eq.~\eqref{dJA} can be matched to the standard stochastic equations for $n_{\rmV}$ and $n_{\rmA}$ in the presence of the CME/CSE~\cite{Iatrakis:2015fma,Lin:2018nxj,Hongo:2018cle}.
Conversely, one might construct an action of a similar form to eq.~\eqref{action} from the stochastic equation following the bottom-up approach of Martin-Siggia-Rose-Janssen-de Dominicis~\cite{PhysRevA.8.423,Janssen,PhysRevB.18.4913}, as was done in ref.~\cite{Hongo:2018cle}.
However, the formalism of refs.~\cite{Crossley:2015,Crossley:2017,Glorioso:2017lcn} as we employ here provides a basis for the systematic analysis.

\subsection{Expansion around thermal equilibrium}
\label{sec:expansion}
Let us consider the fluctuations around the equilibrium state characterized by a static and homogeneous background vector and axial charge densities $(n_{\rmV})_{0}$ and $(n_{\rmA})_{0}$, where the subscript ``$0$'' denotes those equilibrium values. 
In section~\ref{sec:finite-r}, we shall study the situation that axial charge damping coefficient $\g$ is finite so that $(n_{\rmA})_{0}=0$.
In section~\ref{sec:zeror}, 
we take the limit $\g=0$ and consider the systems with a finite $(n_{\rmA})_{0}$.
In both cases, 
we shall use  $\delta n_\alpha= n_\alpha-(n_{\alpha})_0$ as the dynamical fluctuating fields for $\rmr$-variable and $\psi_{\a}$ as the dynamical $\rma$-fields; the latter vanishes in equilibrium.%
\footnote{
Although we have written down the action \eqref{action} explicitly as a functional of $\mu_{\rm V,A}$ to make the dynamical KMS symmetry \eqref{KMS-r} and \eqref{KMS-a} manifest, the resulting vertices will contain time derivatives that would potentially complicate the analysis if we were using $\d\mu_{\rm V,A}$ as the dynamical $\rm r$-variables. 
}
In addition, we rescale $\delta n_\alpha$ and $\psi_{\a}$ for convenience by
\begin{align}
\label{rescale}
\lambda_\alpha = g_\alpha \delta n_\alpha \,,
\quad \psi_\alpha \to g_\alpha \psi_\alpha\,,
\quad
g_\alpha= \frac{1}{\sqrt{T\chi_\alpha}}\, . 
\end{align}
Note that $g^{-2}_{\a}=T\chi_{\a}$ is the equilibrium fluctuation of $\d n_{\a}$ per unit volume. 
This means that the fluctuations of the rescaled variables $\l_{\a}$ are of the order unity, which is real motivation for the definition~\eqref{rescale}. 
Here and throughout this paper, we do {\it not} take the summation over dummy vector/axial indices ($\alpha =\rmV,\rmA$) unless the summation symbol $\Sigma$ is present.

We shall expand the Lagrangian density as
\begin{align}
\sL=(\sL)_{1}+(\sL)_{2}+(\sL)_{3}+\ldots \,,
\end{align}
where the subscript of $(\cdots)$ denotes the number of fluctuating fields.
By construction, $(\sL)_{1}$ is a total derivative. 
We shall use $(\sL)_{2}$ to obtain propagators and read cubic vertices from $(\sL)_{3}$. The quartic vertices from $(\sL)_{4}$ can contribute to one-loop corrections, but such contributions are simply proportional to the UV cut-off and will not be of physical importance. In short, $\sL_{2,3}$ are sufficient for the computing fluctuations corrections at one-loop order.  
Note that if we demand $\sigma_\alpha$, $\gamma$, and $|\Bv|$ to be counted as $g_\alpha^{-2}$, 
then $(\sL)_{n}\sim  g_\alpha^{n-2}$ and $g_{\a}$ can be viewed as the effective coupling constant organizing the fluctuation corrections to the tree-level results. 

To determine $\sL_{2,3}$, we need to expand $\mu_{\rm V,A},\s$, and $\g$ in terms of $\d n_{\rm V,A}$: 
\begin{subequations} 
\label{expand}
\begin{align}
\label{expand-mu}
\mu_{\alpha}&= (\mu_{\alpha})_{1}+(\mu_{\alpha})_{2}+\cdots \,, \\
\s_\alpha &= (\s_{\alpha})_0 + (\s_\alpha)_{1}+\cdots\,,\\ 
\label{expand-gamma}
\g&=(\g)_0 +(\g)_{1}+\cdots\,. 
\end{align}
\end{subequations} 
Defining
\begin{align}
\label{eq:couplings}
&\s_{\alpha;\beta}\equiv 
\frac{1}{\sigma_\alpha} \sqrt{\frac{\chi_\alpha}{\chi_\beta}}
\frac{\pd \s_\alpha}{\pd \mu_{\beta}}\,,\quad
\g_{;\a}\equiv \frac{1}{\gamma} \sqrt{\frac{\chi_\rmA }{\chi_\alpha}}\frac{\pd \g}{\pd \mu_{\alpha}}\,, \quad 
n_{\alpha;\beta\gamma}\equiv \frac{1}{\sqrt{\chi_\beta\chi_\gamma}}\frac{\pd^2 n_{\alpha}}{\pd\mu_{\beta}\pd \mu_{\gamma}}\, , 
\end{align}
we have explicitly
\bes
\label{expand-expression}
\begin{align}
\label{expand-expression-1}
(\mu_\alpha)_{1} &= \sqrt{\frac{T}{\chi_\alpha}} \lambda_\alpha \,,\quad 
(\mu_{\alpha})_{2} = - \frac{T }{2\chi_\alpha} \sum_{\beta,\gamma} n_{\alpha; \beta\g}\lambda_\beta \lambda_\gamma  \,, \\
\label{expand-expression-2}
(\s_\alpha)_1 &= \sigma_\alpha \sqrt{\frac{T}{\chi_\alpha}} \sum_\beta \sigma_{\alpha;\beta} \lambda_\beta \,, \quad
(\g)_{1}= \gamma \sqrt{\frac{T}{\chi_\rmA}} \sum_\alpha \g_{;\alpha} \lambda_\alpha \,.
\end{align}
\ees
Here, the normalizations are chosen to make the counting in terms of $g_{\alpha}$ manifest in the following expressions.
From now on, we omit the label $(...)_0$ for equilibrium quantities when doing so would not lead to confusion. 

Substituting eq.~\eqref{expand} into eq.~\eqref{action} and further using eq.~\eqref{expand-expression}, 
we arrive at the expressions:
\begin{align}
({\sL})_{2}&= \sum_{\alpha} \left[ (\pd_{t}\psi_\alpha) \l_\alpha- D_\alpha (\na \psi_\alpha) \cdot \na \left(\l_\alpha - \i \psi_\alpha \right) \right] + \vv \cdot \left( \l_{\rmV} \na \psi_{\rmA} +  \l_{\rmA} \na \psi_{\rmV} \right) - r \psi_\rmA \left( \l_{\rmA} - \i \psi_\rmA \right)\,, 
\label{eq:L2-scaled}\\
\label{eq:L3}
({\sL})_{3}&= -T \sum_{\alpha,\beta} g_\alpha D_\alpha  \sigma_{\alpha;\beta} \lambda_\beta (\na \psi_\alpha) \cdot \na(\lambda_\alpha -\i \psi_\alpha) 
- \frac{T}{2} \sum_{\alpha,\beta,\gamma} g_\alpha D_\alpha n_{\alpha;\beta \gamma}\l_{\beta}\l_{\gamma} \na^2 \psi_\alpha
\notag\\
& \quad - \frac{T}{2} \sum_{\alpha,\beta} \left(g_\rmV n_{\rmA; \alpha\beta}  \vv_\rmA \cdot \na \psi_\rmV + g_\rmA n_{\rmV; \alpha\beta}  \vv_\rmV \cdot \na \psi_\rmA \right) \lambda_\alpha \lambda_\beta 
\\
& \quad -T g_\rmA r \sum_\alpha  \gamma_{;\alpha}\lambda_\alpha \psi_\rmA \le(\l_{\rm A} - \i \psi_{\rm A}\ri) + \frac{T g_\rmA r}{2} \sum_{\alpha,\beta} n_{\rmA;\alpha\beta} \psi_\rmA \lambda_\alpha \lambda_\beta \notag \,,
\end{align}
with
\begin{align}
D_\alpha \equiv \frac{\sigma_\alpha}{\chi_\alpha}\,, \quad \vv \equiv \frac{C\Bv}{\sqrt{\chi_\rmV \chi_\rmA}} \,,\quad  \vv_\alpha \equiv \frac{C\Bv}{\chi_\alpha} \, , 
\quad
r\equiv \frac{\g}{\chi_{\rm A}}\, . 
\end{align}
Here, $r$ and $\vv$ correspond to the bare axial relation rate and the velocity of the chiral magnetic wave (CMW) \cite{Kharzeev:2010gd,Newman:2005hd}, respectively. 
The first and the third lines in eq.~\eqref{eq:L3} arise from the non-linearity due to the charge diffusion and axial charge relaxation, respectively.  
Since $\mu_{\rm V,A}$ is generically a non-linear function of $n_{\rmV}$ and $n_{\rmA}$, the CME/CSE induce non-linear couplings among fluctuating fields, as is shown in the second line of eq.~\eqref{eq:L3}. 
In the cubic action~\eqref{eq:L3},
the terms involving two a-fields correspond to the multiplicative noise contribution, which is necessary to ensure the KMS invariance.

\subsection{Propagators and vertices}
\label{sec:Rules}

We now define the two-point correlation functions of the fields:
\bes
\label{G-def}
\begin{align}
\CG_{\alpha\beta}^{\rm rr}(x)&=\langle \l_{\alpha}(x)\l_{\beta}(0)\rangle\, , \\
\CG_{\alpha\beta}^{\rm ra}(x) &= \langle \l_{\alpha}(x) \psi_{\beta}(0)\rangle\, ,
\quad
\CG_{\alpha\beta}^{\rm ar}(x)=\langle \psi_{\alpha}(x)\l_{\beta}(0)\rangle\,,
\end{align}
\ees
while $\langle \psi_{\alpha}(x)\psi_{\beta}(0)\rangle=0$ by causality.

\begin{figure}[t]
\begin{center}
\includegraphics[bb=0 0 850 220, width=12cm]{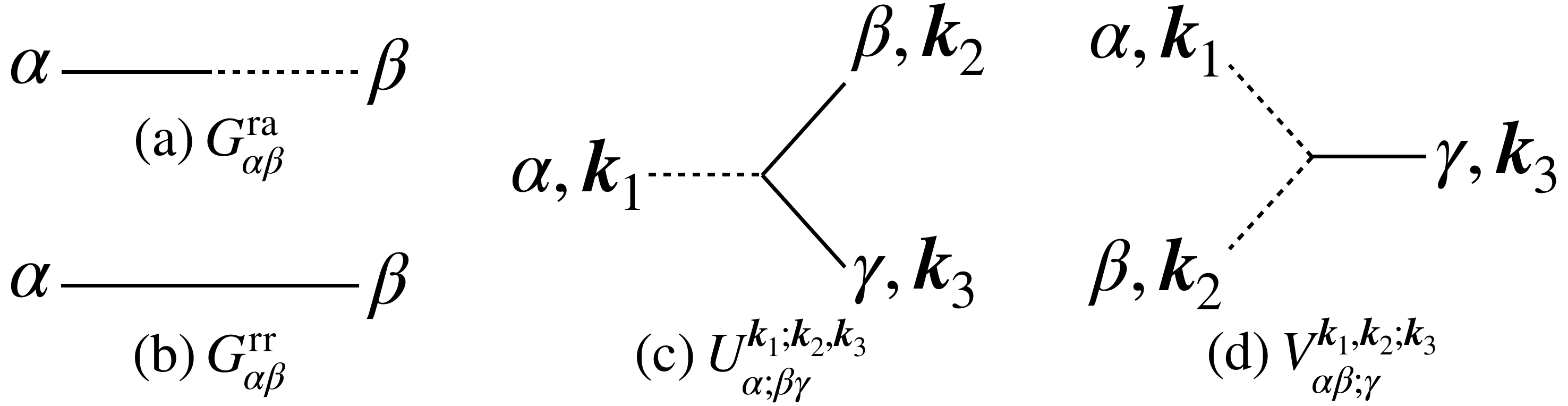}
\caption{
Diagrammatic representations of (a) $G^{\rmr \rma}_{\alpha\beta}$ , (b) $G^{\rmr \rmr}_{\alpha\beta}$, and the vertices (c) $U$ and  (d) $V$. 
The solid and dotted lines correspond to $\l$ (r-field) and $\psi$ (a-field), respectively. }
\label{Feynman-rules}
\end{center}
\end{figure}

To perform the diagrammatic analysis, 
we shall consider the free propagators $G^{\rm rr}_{\alpha\beta}, G^{\rm ra}_{\alpha\beta}$, and $G^{\rm ar}_{\alpha\beta}$, which are $\mathcal G^{\rm rr}_{\alpha\beta}, \mathcal G^{\rm ra}_{\alpha\beta}$, and $\mathcal G^{\rm ar}_{\alpha\beta}$ at the tree level, respectively. 
Suppressing the indices $\alpha$ and $\beta$, we can read their expressions from $(\sL)_2$ given by eq.~\eqref{eq:L2-scaled} as
\begin{eqnarray}
\label{G-inv}
\left(
\begin{array}{cc}
G^{\rm rr} & G^{\rm ra} \\
G^{\rm ar} & 0
\end{array}
\right)&=&
\left(
\begin{array}{cc}
0 & -\i M^\dagger \,\\
-\i M & N 
\end{array}
\right)^{-1}\, ,
\end{eqnarray}
where in the Fourier space with $K^\mu = (\omega,\kv)$,
\begin{eqnarray}
\label{BC-def}
M (K) =
\left(
\begin{array}{cc}
\i \omega - D_\rmV \kv^{2}
& -\i \vv \cdot \kv \\
-\i \vv \cdot \kv & \i \omega - (r+D_\rmA \kv^{2})
\end{array}
\right)\,,
\quad
N(\kv) =  2 \left(
\begin{array}{cc}
\displaystyle  D_\rmV \kv^{2} & 0\\ 
0 & r+D_\rmA \kv^{2}
\end{array}
\right)\, .
\end{eqnarray}
Note that $M^{\rm T} = M$. 
From eq.~\eqref{G-inv}, we obtain
\begin{eqnarray}
\label{G-BC}
G^{\rm ra}=\i M^{-1}\, , 
\quad
G^{\rm ar}=\i (M^{-1})^\dagger\, , 
\quad
G^{\rm rr}=-G^{\rm ra} N G^{\rm ar}\, .
\end{eqnarray}
The diagrammatic representations of $G^{\rmr \rma}_{\alpha\beta}$ and $G^{\rmr \rmr}_{\alpha\beta}$ are given by figures~\ref{Feynman-rules} (a) and (b), respectively. 

The retarded propagator $G^{{\rm ra}}$ is the basic building block in the subsequent diagrammatic computations, whereas $G^{\rm ar}$ and $G^{\rm rr}$ can be expressed in terms of $G^{\rm ra}$:
\begin{align}
\label{eq:Gra-Gar}
G^{\rma\rmr}(K) & =-(G^{\rmr\rma})^{\dagger}(K)=G^{\rmr\rma}(-K) \, ,
\\
\label{eq:FDT2}
G^{\rmr \rmr}(K) &=\i \left[G^{\rmr \rma}(K) + G^{\rma \rmr}(K) \right]\,,
\end{align}
as one can verify explicitly from eqs.~\eqref{G-BC} and \eqref{BC-def}. 
A particular useful form for $G^{{\rm ra}}$ is that in a Laurent expansion:
\begin{align}
\label{Laurent}
G ^{\rm ra}(K)=\sum_{m=\pm } \frac{R^{m}(\vk)}{\o-\Omega^{m}(\vk)}\, ,
\end{align}
where $m=+,-$ labels two independent collective modes with
\begin{align}
\label{O-tree}
\Omega^{\pm}(\kv)&=-\frac{\i}{2}(r+ D_\rmV \kv^{2}+D_\rmA \kv^{2})\pm \frac{1}{2} \sqrt{4(\vv \cdot\vk)^{2}-(r+D_\rmA \kv^2 - D_\rmV \kv^2 )^{2}}\,, 
\\
\label{R-pole}
R^{\pm}(\vk)&=
\frac{\pm 1}{\Omega_{+}(\vk)-\Omega_{-}(\vk)}\left(
\begin{array}{cc}
\Omega_\pm(\vk) + \i(r + D_\rmA \kv^2) &{\bm v} \cdot \kv \\
{\bm v} \cdot \kv & \Omega_\pm(\vk) + \i D_\rmV \kv^2
\end{array}
\right) \,.
\end{align}
In the limit $r=0$ and for $\vk\cdot\hat{\Bv}>0$, these two modes $(m=+,-)$ correspond to the CMW propagating in the same/opposite directions to the magnetic field, respectively. 

Next, we define the interaction vertices from $\sL_{3}$ as
\begin{align}
 \i\int_x  (\mL)_{3} &= T \sum_{\alpha,\beta,\gamma}g_\alpha \int \dd t \int_{\kv_1,\kv_2,\kv_3}\le(
 U_{\alpha \beta \gamma}^{\kv_1;\kv_2,\kv_3} \psi_{\alpha}^{\kv_1} \l_{\beta}^{\kv_2} \l_{\gamma}^{\kv_3} + V_{\alpha \beta \gamma}^{\kv_1,\kv_2;\kv_3} \psi_\alpha^{\kv_1} \psi_\beta^{\kv_2} \l_\gamma^{\kv_3}
 \ri) \, ,
\end{align}
where $\psi^{\kv}_{\a}\equiv\psi(t,\kv)$ and $\l^{\kv}_{\a}\equiv\l(t,\kv)$. 
There are two types of vertices: $U$ couples one a-field with two r-fields; $V$ couples two a-fields with one r-field. They can be read from the cubic action~\eqref{eq:L3}
\begin{subequations} 
\label{UVs}
\begin{align}
\label{eq:tilde-V-raa-Vij}
 U_{\rmV;\alpha\beta}^{\kv_1;\kv_2,\kv_3} &= 
\frac{-\i D_\rmV}{2} \left(
\begin{array}{cc}
 \sigma_{\rmV;\rmV}  \kv^2_1 &-  \sigma_{\rmV;\rmA}   \kv_1\cdot\kv_2\\
-  \sigma_{\rmV;\rmA} \kv_1\cdot\kv_3 & 0 
\end{array}
\right)
+\frac{\i}{2} D_\rmV  n_{\rmV;\alpha\beta} \kv^2_1 +\frac{1}{2} \vv_\rmA \cdot \kv_1  n_{\rmA;\alpha\beta} \,,
\\
\label{UA}
U_{\rmA ; \alpha\beta}^{\kv_1;\kv_2,\kv_3} &=
\frac{-\i D_\rmA}{2}\left(
\begin{array}{cc}
0 & -  \sigma_{\rmA;\rmV}  \kv_1\cdot\kv_3 \\
-  \sigma_{\rmA;\rmV}  \kv_1\cdot \kv_2 &  \sigma_{\rmA;\rmA} \kv_1^2
\end{array}
\right)
+ \frac{\i}{2} D_\rmA  n_{\rmA;\alpha\beta}\kv_1^2 + \frac{1}{2}\vv_\rmV \cdot \kv_1  n_{\rmV;\alpha\beta}\notag\\
&\quad - \frac{\i r }{2} \left(
\begin{array}{cc}
0 &  \gamma_{;\rmV}  \\
 \gamma_{;\rmV}  & 2  \gamma_{;\rmA}
\end{array}
\right) +\frac{\i r}{2}  n_{\rmA;\alpha\beta} \,, \\
V_{\rmV \alpha; \beta}^{\kv_1,\kv_2;\kv_3} &=D_\rmV \kv_1\cdot \kv_2
\left(
\begin{array}{cc}
 \sigma_{\rmV;\rmV} & \sigma_{\rmV;\rmA}  \\
0 & 0
\end{array}
\right) \,, \\
\label{aarA}
V_{\rmA \alpha; \beta}^{\kv_1,\kv_2;\kv_3} &= 
D_\rmA \kv_1\cdot \kv_2
\left(
\begin{array}{cc}
0 & 0 \\
 \sigma_{\rmA;\rmV} & \sigma_{\rmA;\rmA} 
\end{array}
\right) 
- r \left(
\begin{array}{cc}
0&0 \\
 \gamma_{;\rmV}  &   \gamma_{;\rmA} 
\end{array}
\right) \,.
\end{align}
\end{subequations}
Note, the first term in the second line of eq.~\eqref{UA} and the last term in eq.~\eqref{aarA} arise from the fact that the axial damping coefficient $\gamma$ depends on $n_{\rm V}$ and $n_{\rm A}$.

The graphic presentations of $U^{\kv_1,\kv_2,\kv_3}_{\alpha\beta\gamma}$ and $V^{\kv_1,\kv_2,\kv_3}_{\alpha\beta\gamma}$ are shown in figures~\ref{Feynman-rules} (c) and (d), respectively. 
With the propagators and vertices at hand, we are ready to compute one-loop corrections to the conductivity.

\section{Conductivity at finite axial relaxation rate}
\label{sec:con}

\subsection{Conductivity}
\label{sec:Kubo}

From the symmetrized correlator of the vector current $J^{\mu}$, $\CC^{\mu\nu}\sim \langle J^{\mu} J^{\nu}\rangle$, we can determine the (AC) conductivity tensor through the standard Kubo formula
\begin{align}
\label{eq:Kubo-0}
\sigma^{ij} (\omega) = \lim_{\kv \rightarrow \zev} \frac{1}{2T} \mathcal C^{ij} (K)\,.
\end{align}
Alternatively, we can extract $\s^{ij}$ from the retarded correlator (see eq.~\eqref{Kubo-R} in section \ref{sec:zeror}). 
The conductivity tensor in the presence of the external magnetic field can be decomposed as 
\begin{align}
\label{eq:sigma-decomp}
\sigma^{ij} = \sigma_{\parallel}\hat B^i \hat B^j + \sigma_{\perp} (\delta^{ij}-\hat B^i\hat B^j)  \,,
\end{align}
where $\s_{\parallel}$ and $\s_{\perp}$ are the longitudinal and transverse conductivity, respectively. In this work, we will not consider the Hall conductivity. 
From the Ward-Takahashi identity, 
we also have
\begin{align}
\label{eq:Kubo-2}
\lim_{\kv \rightarrow \zev} \frac{1}{2T}\frac{\omega^2}{\kv^2} \mathcal \CC^{00}(K) =
\lim_{\kv \rightarrow \zev} \frac{1}{2T} \hat k^i \hat k^j \mathcal C^{ij} (K) = \sigma_{\perp} (\omega) +
[\sigma_{\parallel}(\omega)-\sigma_{\perp}(\omega)]
 (\hat\kv \cdot \hat\Bv)^2\,, 
\end{align}
which allows us to extract $\s_{\parallel}$ and $\s_{\perp}$ from the small $k$ behavior of $\CC^{00}$.

In what follows, we determine $\CC^{00}$ and hence $\s^{ij}(\o)$ from the relation,
\begin{align}
\label{eq:Cg}
\CC^{00}=  T \chi_{\rm V} \CG_{\rm VV}^{\rm rr}\,, 
\end{align}
which follows from the definitions of the rescaled fields, eq.~\eqref{rescale}. At tree level, 
an explicit calculation using eq.~\eqref{G-BC} yields
\begin{align}
\label{G00-tree}
\lim_{\vk\to \zev} \frac{\o^{2}}{2\kv^{2}}G^{\rm rr}_{\rm VV}(K)= 
D_{\rm V} + \frac{{\bm v}^{2} r}{\o^{2}+r^{2}}  \,. 
\end{align}
It then follows from eqs.~\eqref{eq:Kubo-2} and \eqref{eq:sigma-decomp} that
\begin{align}
\label{sigma-tree}
(\s_{\parallel})_{{\rm tree}}(\o)= (\s_{\rm V})_{0} + \frac{C^2 {\bm B}^2 r}{\chi_{\rm A}(r^{2}+\o^{2})}\,, 
\quad
(\s_{\perp})_{{\rm tree}}(\o)= (\s_{\rm V})_{0} \,,
\end{align}
which reproduces the well-known CME-induced negative magnetoresistance~\cite{Son:2012bg}. However, the tree level result~\eqref{sigma-tree} does not take into account the fluctuation effects. We shall study the one-loop corrections to $\s^{ij}$ by computing $\CG^{\rm rr}_{\rm VV}$ dressed by the self-energies.

\subsection{Self-energies}
\label{sec:relation}
\begin{figure}[t]
\begin{center}
\includegraphics[bb=0 0 900 210, width=15cm]{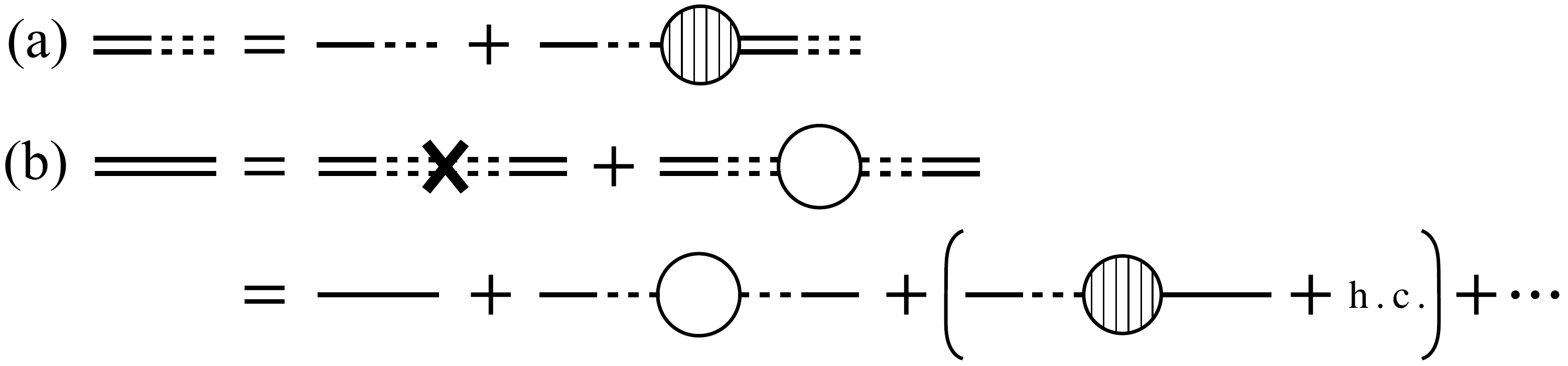}
\caption{Diagrammatic representations of the Dyson equations: (a) for the $\mathcal G^{\rmr \rma}$ propagator (\ref{eq:Dyson-Gra}); (b) for the $\mathcal G^{\rmr \rmr}$ propagator (\ref{eq:Dyson-Grr}). The last expression on (b) is the first-order expansion with respect to the self-energies. Here, double lines represent the full propagators. The white bubbles and the bubble with diagonal lines correspond to $\Sigma^{\rma \rma}$ and $\Sigma^{\rma \rmr}$, respectively. The cross vertex denotes $-N$, which is related to the free propagators by eq.~(\ref{G-BC}). Note that $(\alpha,\beta)$ labels are suppressed in these diagrams.}
\label{fig:Dyson}
\end{center}
\end{figure}

We start from the full propagators which are dressed by self-energies $\Sigma^{\rm ar}, \Sigma^{\rm ra}$, and $ \Sigma^{\rm aa}$ through the Dyson equation:
\begin{eqnarray}
\label{DS}
\left(
\begin{array}{cc}
 \CG^{\rm rr} & \CG^{\rm ra} \\
\CG^{\rm ar} & 0
\end{array}
\right)^{-1}
&=&
\left(
\begin{array}{cc}
G^{\rm rr} & G^{\rm ra}\\
G^{\rm ar} & 0
\end{array}
\right)^{-1}
-
\left(
\begin{array}{cc}
0 & \Sigma^{\rmr \rma} \\
\Sigma^{\rma \rmr} & \Sigma^{\rmr \rmr}
\end{array}
\right)\,,
\end{eqnarray}
or equivalently,
\begin{align}
\label{eq:Dyson-Gra}
{\cal G}^{\rmr \rma} &= [(G^{\rmr\rma})^{-1}  - \Sigma^{\rma\rmr} ]^{-1}=(- \i M  - \Sigma^{\rma\rmr} )^{-1} \,, \\ 
\label{eq:Dyson-Grr}
{\cal G}^{\rmr \rmr} & =- \CG^{\rmr\rma} (N-\Sigma^{\rma\rma}) \CG^{\rm ar}  \,. 
\end{align}
See figures~\ref{fig:Dyson} (a) and (b) for graphical representations of eqs.~(\ref{eq:Dyson-Gra}) and (\ref{eq:Dyson-Grr}), respectively. 

To determine the small $k$ behavior of $\CG^{\rm rr}_{\rm VV}$ from eq.~\eqref{eq:Dyson-Grr},
we shall first consider the behavior of the self-energies in this limit. We note $\Sigma^{\rm aa}_{\a\b}$ can be connected to two external legs associated with  $\psi_{\a}$ and $\psi_{\b}$. 
Since $\psi_{\rm V}$ is always combined with $\na$ in the cubic Lagrangian density $(\sL)_{3}$ in eq.~\eqref{eq:L3}, we have $\Sigma^{\rm aa}_{\a\b}\sim k^{\delta_{\a \rmV}+\delta_{\b \rmV}}$ and $\Sigma^{\rm ar}_{\a\b}\sim k^{\delta_{\a \rmV}}$, where $\delta_{\alpha \beta}$ is the Kronecker delta. 
Then, the relevant components of the self-energies can be parametrized as
\bes
\label{Delta-def}
\begin{align}
\label{Delta-def-Sigma-aa-VV}
(\chi_\rmV)_0 \Sigma^{\rma\rma}_{\rmV\rmV} (K) &= - 2\Delta \sigma_{\perp} (\omega) \kv^2 -2 \left[
\Delta \sigma_{\parallel}(\omega)  -\Delta \sigma_{\perp}(\omega) \right](\kv \cdot \hat\Bv)^2 +\ldots\,, 
\\
\label{Delta-gamma}
(\chi_\rmA)_0 \Sigma^{\rma\rma}_{\rm AA}(K) &= -2\Delta \gamma(\o)+\ldots\,, \\
\label{Delta-r}
\Sigma^{\rma\rmr}_{\rm AA}(K) &= -\i \Delta r(\o)+\ldots\, , \\
\label{Delta-def-end}
\Sigma_{\rmV\rmA}^{\rma\rmr}(K) &= \kv \cdot \Delta \vv_{\rm A} (\o)+\ldots\, ,
\end{align}
\ees
where $\ldots$ are the terms suppressed by small $k$. Note that $\Delta \g, \Delta r$, and $\Delta \vv_{\rm A}$ can be viewed as the finite frequency corrections to $\g, r$, and $\vv$, respectively. 
To see this, one should keep in mind that $\Sigma^{\rm aa}$ and $\Sigma^{\rm ar}$ enter as the correction to $-N$ and $\i M$, respectively, in eqs.~\eqref{eq:Dyson-Grr} and \eqref{eq:Dyson-Gra} while $ (\chi_{\rm A})_{0} N_{\rm AA}=2\g+{\cal O}(k^{2})\,, M_{\rm AA}=-r+{\cal O}(k^2)$, and $M_{\rm VA}=-{\rm i} (\vk\cdot\vv) + \mathcal O(k^2)$.

By substituting eq.~\eqref{Delta-def} into the Dyson equation \eqref{eq:Dyson-Grr} and evaluate the Kubo relation \eqref{eq:Kubo-2} with \eqref{eq:Cg}, we eventually find 
\bes
\label{eq:Sigma-expr}
\begin{align}
\label{eq:Sigma-iso-expr}
\sigma_{\perp}(\omega) & = (\sigma_{\rmV})_0 + 
\Delta \sigma_{\perp} (\omega) \,, \\ 
\label{eq:Sigma-an-expr}
\sigma_{\parallel} (\omega) & =
(\sigma_{\rmV})_0 
+
 \Delta\sigma_{\parallel} (\omega) + \frac{[(\vv)_0 + \Delta \vv_\rmA (\omega)]^{2} [ (\chi_\rmA r)_0 +\Delta \gamma(\omega)](\chi_\rmV)_0}{(\chi_\rmA)_0 \le[\omega^{2}+\le((r)_0+\Delta r(\omega)\ri)^{2}\ri]} \, ,
\end{align}
\ees
meaning the loop corrections to the conductivity can be expressed in terms of the following $\omega$-dependent functions: 
\begin{align}
\label{RG-list}
\Delta \sigma_{\perp} (\o)\,,\quad 
\Delta \sigma_{\parallel} (\o)\,,\quad \Delta \gamma(\o)\,, \quad \Delta r(\o)\,, \quad \Delta \vv_{\rm A}(\o)\,, 
\end{align}
which we shall compute in section \ref{sec:finite-r}. 
Intuitively, we may understand eq.~(\ref{eq:Sigma-expr}) by replacing $(\sigma_\rmV)_0$, $\gamma$, $r$, and $\vv$ of eq.~(\ref{sigma-tree}) into those including fluctuation corrections \eqref{RG-list}.

\subsection{One-loop}
\label{sec:one-loop}
\begin{figure}[t]
\begin{center}
\includegraphics[bb=0 0 1800 760, width=15cm]{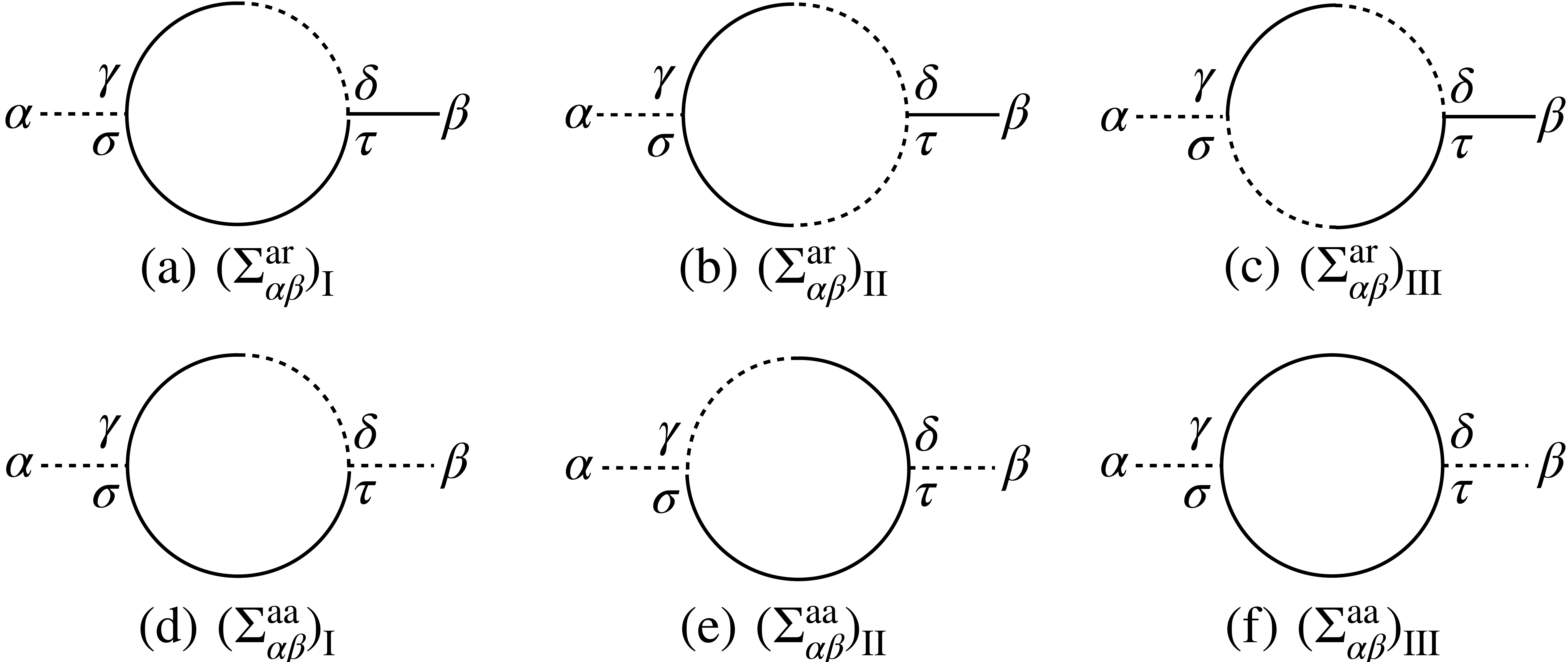}
\caption{Self-energies at the one-loop order.}
\label{fig:Sigmas}
\end{center}
\end{figure}

In this subsection, we provide general expressions for computing the self-energies $\Sigma^{\rma\rmr}_{\alpha\beta}$ and $\Sigma^{\rma\rma}_{\alpha\beta}$ at the one-loop level. 
We will give the derivation for the former and present only the results for the latter, which can be derived similarly (see appendix~\ref{sec:self-aa} for details).

The self-energy $\Sigma^{\rma\rmr}_{\alpha\beta}$ consists of three pieces:
\begin{align}
\label{eq:Sigma-ar-alphabeta-figs}
\Sigma_{\alpha\beta}^{\rma\rmr} =(\Sigma_{\alpha\beta}^{\rma\rmr})_{\rm I} +(\Sigma_{\alpha\beta}^{\rma\rmr})_{\rm II} +(\Sigma_{\alpha\beta}^{\rma\rmr})_{\rm III} \,,
\end{align}
whose diagrammatic representations are  given by figures~\ref{fig:Sigmas} (a)--(c), respectively. First of all, the contribution from $(\Sigma_{\alpha\beta}^{\rma\rmr})_{\rm III}$ in figure~\ref{fig:Sigmas} (c) vanishes. 
This is because
\begin{align}
\label{dq0-zero}
\int_{q^0} G_{\alpha\beta}^{\rm ra}(q^0,\vq_{+}) G_{\gamma\delta}^{\rm ar}(\o-q^0,-\vq_{-})=0 \,,
\end{align}
where the integrand has poles only in the upper complex $q^0$-plane so that the contour integral vanishes.
Here and hereafter, we use the notation $\vq_{\pm}=\vq\pm \vk/2$.

For the first term in eq.~(\ref{eq:Sigma-ar-alphabeta-figs}), we have 
\begin{align}
\label{Sigma-ar-I}
(\Sigma_{\alpha\beta}^{\rma\rmr})_{\rm I} (K)
&= 4 T^2 \sum_{\gamma\delta\sigma\tau}\int_{Q} g_\alpha U_{\alpha;\gamma\sigma}^{-\kv;-\vq_{-},\vq_{+}} G^{{\rm rr}}_{\sigma\tau}(q^0,\vq_{+}) G^{{\rm ra}}_{\gamma \delta}(\o-q^0,-\vq_{-}) g_\delta
U_{\delta;\beta \tau}^{\vq_{-};\kv,-\vq_{+}} 
 \no \\
&=4 \i T^{2} \sum_{\gamma\delta\sigma\tau} \int_{Q} g_\alpha U_{\alpha;\gamma\sigma}^{-\kv;-\vq_{-},\vq_{+}} G^{{\rm ra}}_{\sigma\tau}(q^0,\vq_{+}) G^{{\rm ra}}_{\gamma \delta}(\o-q^0,-\vq_{-}) g_\delta
U_{\delta;\beta\tau}^{\vq_{-};\kv,-\vq_{+}}  \,.
\end{align}
To obtain the second line from the first line, we have replaced $G^{\rm rr}(q^0,\qv_+)$ with $\i G^{\rm ra}(q^0,\qv_+)$ using eq.~\eqref{eq:FDT2} together with eq.~\eqref{dq0-zero}.  Turning to $(\Sigma^{\rma\rmr}_{\alpha\beta})_{\rm II}$, we have
\begin{align}
(\Sigma^{\rma\rmr}_{\alpha\beta})_{\rm II} (K)  
&= 2 T^{2}\sum_{\gamma\delta\sigma\tau}
\int_{Q} 
g_\alpha U_{\alpha;\gamma\sigma}^{-\kv;-\qv_{-},\qv_{+}} G^{{\rm ra}}_{\sigma\tau}(q^0,\vq_{+}) G^{{\rm ra}}_{\gamma\delta}(\o-q^0,-\vq_{-}) g_\delta V_{\delta\tau;\beta}^{\vq_{-},-\vq_{+};\kv}\,.
\end{align}
Adding $(\Sigma^{\rma\rmr}_{\alpha\beta})_{\rm I}$ and $(\Sigma^{\rma\rmr}_{\alpha\beta})_{\rm II}$, the self-energy becomes
\begin{align}
\label{Sigma-ar-final0}
\Sigma^{\rma\rmr}_{\alpha\beta}(K)=T^{2}\int_{Q} {\rm tr} \le[G^{\rm ra}(q^0,\vq_{+})
(\Gamma_\alpha^\rma)^{\rm T} G^{\rm ra}(\o-q^0,-\vq_{-}) \Gamma_\beta^\rmr \ri]\, , 
\end{align}
where we have defined two-by-two matrices: 
\begin{align}
\label{Gamma-1}
(\Gamma^\rma_\alpha)_{\beta\gamma} = g_\alpha U_{\alpha;\beta\gamma}^{-\vk;-\vq_{-},\vq_{+}}\, , \quad(\Gamma^\rmr_\beta)_{\gamma\delta} = 4 \i g_\gamma U_{\gamma;\beta\delta}^{\qv_- ; \kv,-\qv_+}+2 g_\gamma V_{\gamma\delta;\beta}^{\vq_{-},-\qv_+;\kv}\,.
\end{align}

We now carry out the integration over $q^0$ in eq.~(\ref{Sigma-ar-final0}) using the Cauchy residue theorem, yielding
\begin{align}
\label{q0-int}
\int_{q^0} G^{{\rm ra}}_{\sigma\tau}(q^0,\vq_{+}) G^{{\rm ra}}_{\gamma \delta}(\o-q^0,-\vq_{-})=-\i\sum_{m,m'} R^m_{\sigma\tau}(\vq_{+})R^{m'}_{\gamma\delta}(-\vq_{-}) \Delta^{mm'}(K,\vq)\,,
\end{align}
where we have used eq.~(\ref{Laurent}) and introduced the notation,
\begin{align}
\label{eq:eff-prop}
\Delta^{mm'}(K,\vq)\equiv\frac{1}{\o-\Omega^{m}(\vq_{+})-\Omega^{m'}(-\vq_{-})}\,.
\end{align}
Substituting eq.~\eqref{q0-int} into eq.~\eqref{Sigma-ar-final0} leads to the final result
\begin{align}
\label{Sigma-ar-final}
& \Sigma^{\rma \rmr}_{\alpha\beta}(K)= -\i T^2 \sum_{m,m'}\int_{\vq} F^{\rma \rmr,mm'}_{\alpha\beta}(\kv,\qv)
\Delta^{m m'}(K,\vq)
\,,
\end{align}
where we have defined
\begin{align}
\label{tF}
F^{\rma \rmr,mm'}_{\alpha\beta}(\kv,\qv) = {\rm tr} \le[
R^m (\vq_{+}) (\Gamma_\alpha^\rma)^{\rm T} R^{m'}(-\vq_{-}) \Gamma^\rmr_\beta \ri]\,.
\end{align}

One can also analyze $\Sigma^{\rm aa}_{\alpha\beta}$ following the similar treatment (see appendix~\ref{sec:self-aa} for details). 
For our purpose, only the diagonal components of $\Sigma^{\rm aa}_{\a\b}$ are needed; they read 
\begin{align}
\label{Sigma-aa-final}
\Sigma^{\rma \rma}_{\alpha\a}(K)&= T^2\,{\rm Im}\sum_{m,m'}\int_{\vq} 
F^{\rma \rma,mm'}_{\alpha\a}(\kv,\qv) \Delta^{mm'}(K,\vq) \,,
\end{align}
where we have defined
\begin{align}
\label{F}
F^{\rma \rma,mm'}_{\alpha\beta}(\kv,\qv)={\rm tr} \le[ R^m 
(\vq_{+})(\Gamma^\rma_\alpha)^{\rm T}R^{m'}(-\vq_{-}) \Gamma^{'\rma}_\beta \ri]\,,
\end{align}
with 
\begin{align}
\label{Gamma-2}
(\Gamma^{'\rma}_\beta)_{\gamma\delta}=4\le(-g_\beta U_{\beta;\gamma\delta}^{\vk;\vq_{-},- \vq_{+}}+2 \i g_\gamma V_{\gamma\beta;\delta}^{\vq_{-},\vk;-\vq_{+}}\ri) \,.
\end{align}

Expressions~\eqref{Sigma-ar-final} for $\Sigma^{\rma\rmr}_{\alpha\beta}$ and \eqref{Sigma-aa-final} for $\Sigma^{\rma\rma}_{\alpha\alpha}$ are the main results of this subsection.
These are to be used in the following subsection.

\subsection{Results}
\label{sec:finite-r}

We now compute one-loop contributions to the conductivity tensor. 
We first note that in the absence of the background axial charge density,
\begin{align}
n_{\rmA;\rmA\rmA}=
n_{\rmV;\rmV\rmA}=n_{\rmV;\rmA\rmV}=n_{\rmA;\rmV\rmV}=0,\quad \sigma_{\rmV;\rmA}=\sigma_{\rmA;\rmA}=0,\quad \gamma_{;\rmA} =0\,.
\end{align}
Let us further assume, for the present illustrative purpose, 
\begin{align}
\label{eq:chi-sigma-non-mixing}
\chi_\rmV = \chi_\rmA = \chi,\quad \sigma_\rmV = \sigma_\rmA = \sigma\,.
\end{align}
Then, the vertices are parameterized by three independent parameters:
\begin{align}
u_n \equiv n_{\rmV;\rmV\rmV} = n_{\rmV;\rmA\rmA} = n_{\rmA;\rmV\rmA}= n_{\rmA;\rmA\rmV}\,,\quad u_\sigma \equiv \sigma_{\rmV;\rmV}=\sigma_{\rmA;\rmV}\,, \quad u_\gamma \equiv \gamma_{;\rmV}\,.
\end{align}
In this subsection, we will drop the subscript in $D,\vv$, and $g$ because of eq.~\eqref{eq:chi-sigma-non-mixing}.  

To evaluate eqs.~\eqref{Sigma-ar-final} and \eqref{Sigma-aa-final}, 
we first consider relevant
$\G^\rma_\alpha,\G^{'\rma}_\alpha,$ and $\G^\rmr_\alpha$, which, in the small $k$ limit, 
are simplified as
\bes
\label{Gamma-small-k}
\begin{align}
\label{Gamma1-VA}
\Gamma_\rmV^\rma &= -\frac{g}{2}(\vv \cdot\kv) \left(
\begin{array}{cc}
0 & u_n \\
u_n & 0
\end{array}
\right) + \mathcal O(k^2)\,, \\
\Gamma_\rmA^\rma  &=
 \frac{\i g r}{2} \left(
\begin{array}{cc}
0 & u_n - u_\gamma \\
 u_n - u_\gamma & 0 
\end{array}
\right)+ \mathcal O(k) \,,
\\
\G_{\rm A}^{'\rma}&=
 -2 \i g r \left(
\begin{array}{cc}
0 & u_n -u_\gamma \\
u_n + 3 u_\gamma & 0 
\end{array}
\right)+\mathcal O(k) \,,
\\
\G_{\rm V}^{'\rma}&= 
8 \i g D (\vk \cdot \qv)
\left(
\begin{array}{cc}
u_\sigma  &0  \\
0 & 0
\end{array}
\right) - 2 g (\vv \cdot \vk) \left(
\begin{array}{cc}
0 & u_n \\
u_n & 0
\end{array}
\right) + \mathcal O(k^2) \,,
\\
\G_{\rm A}^{\rmr}&=
 - 2g\qv^2 D \left(
\begin{array}{cc}
0 & u_n \\
 u_n & 0 
\end{array}
\right) + 2 \i g (\vv \cdot \qv) \left(
\begin{array}{cc}
 u_n & 0 \\
0 & u_n 
\end{array}
\right)
+ 2gr \left(
\begin{array}{cc}
0 & 0 \\
u_\gamma - u_n & 0 
\end{array}
\right)+{\cal O}(k) \,.
\end{align}
\ees
Applying eq.~\eqref{Gamma-small-k} to $F^{\rma\rma,mm'}_{\alpha\beta}$ and $F^{\rma\rmr,mm'}_{\alpha\beta}$, defined in eqs.~\eqref{F} and \eqref{tF}, respectively, 
one can directly verify the parametric behaviors:
\bes
\label{F-same-sign}
\begin{align}
F_{\rmV\rmV}^{\rma\rma} &\sim k^2 \,,
\quad
F_{\rmV\rmA}^{\rma\rmr} \sim k \,,
\quad
F_{\rmA\rmA}^{\rma\rma} \sim F_{\rmA\rmA}^{\rma\rmr} \sim 
k^0\, \quad \ (m \neq m')\,,
\\
F_{\rmV\rmV}^{\rma\rma}&\sim k^3 \,, 
\quad
F_{\rmV\rmA}^{\rma\rmr} \sim k^2\,, 
\quad
F_{\rmA\rmA}^{\rma\rma} \sim F_{\rmA\rmA}^{\rma\rmr} \sim k\,
\quad\ (m = m')\,.
\end{align}
\ees
Comparing eq.~\eqref{F-same-sign} with eq.~\eqref{Delta-def}, 
we observe that only the contribution from $m\neq m'$ are needed to evaluate the functions listed in eq.~(\ref{RG-list}). Therefore, we can replace $\Delta^{mm'}$ in eqs.~\eqref{Sigma-ar-final} and \eqref{Sigma-aa-final} with their expression in the limit $\kv \rightarrow\zev$:
\begin{align}
\label{eq:Delta-pm-expr}
\lim_{\kv\to \zev}\Delta^{+-}(K,\vq)
=\lim_{\kv\to\zev}\Delta^{-+}(K,\vq)=\frac{1}{\o+\i r+2\i D\qv^2}\,.
\end{align}

We first evaluate $\Sigma^{\rma\rma}_{\rm VV}$ which in turn determines $\Delta \sigma_{\parallel}(\omega)$ and $\Delta \sigma_{\perp}(\omega)$. 
From eqs.~(\ref{Gamma-small-k}) and \eqref{F}, 
we have
\begin{align}
\label{F-VV}
F_{\rm VV}^{\rma\rma,+-}+F_{\rm VV}^{\rma\rma,-+}= 2 g^2 u_n^2  ({\bm v}\cdot \vk)^{2}+{\cal O}(k^3)\,,
\end{align}
so that eq.~\eqref{Sigma-aa-final} becomes
\begin{align}
\label{Sigma-VV-0}
\Sigma_{\rm VV}^{\rma\rma}(K)= 2 g^{2} T^2 u_n^2 ({\bm v}\cdot \vk)^{2} \, {\rm Im} \int_{\vq} \frac{1}{\o+\i r+2\i D\qv^2} +{\cal O}(k^4)\, . 
\end{align}
Comparing eq.~\eqref{Sigma-VV-0} with eq.~\eqref{Delta-def-Sigma-aa-VV}, 
we find
\begin{align}
\label{Delta-D-omega}
\Delta \sigma_{\parallel} (\o) =- \frac{g^2 T^2 \chi u_n^2 \vv^2 }{8 \pi D}\,
{\rm Re} \sqrt{\frac{r-\i\o }{2D}} \, ,
\quad
\Delta \sigma_{\perp}(\omega)=0\,.
\end{align}
Here and below, we use the dimensional regularization to perform the integration over $\qv$,
\begin{align}
\label{DR}
\int_\qv \frac{1}{\omega+\i r+ 2\i D \qv^2} = \frac{\i }{8\pi D } \sqrt{\frac{r-\i \omega}{2D}} \,,
\quad
\,\int_\qv \frac{ \qv^2}{\omega+\i r+ 2\i D \qv^2} = \frac{-\i }{8 \pi D } \left(\frac{r-\i \omega}{2D}\right)^{\frac{3}{2}}\,. 
\end{align}
Evaluating loop integrals using dimensional regularization is essentially picking up the contribution from the IR scale, which only depends on the physical inputs but does not depend on the unphysical UV cut-off $\Lambda$ of the EFT (see, e.g., ref.~\cite{Manohar:2018aog}). 
For instance, in eqs.~\eqref{DR}, the IR momentum scale is given by
\begin{align}
\label{eq:qr*}
q^{*}_{r}(\o)=\left| \sqrt{\frac{r-\i \o}{2D}} \right|\,,
\end{align}
which arises from the competition between axial relaxation and diffusion at finite $r$. 
Such IR momentum scale can be interpreted as the characteristic momentum of fluctuation modes and has to be parametrically smaller than the cut-off scale of the EFT, which, in the present case, is $\Lambda\sim l^{-1}_{{\rm mfp}}$. Here, $l_{{\rm mfp}}$ denotes the mean free path.
Indeed, from $D \sim  l_{{\rm mfp}}$, it is easy to verify that $q^{*}_{r}\ll \Lambda$. 
Note that at $r = 0$, $q^{*}_{r}\sim \sqrt{\o/D}$, which leads to hydrodynamic long-time tail behavior \cite{POMEAU197563,Kovtun:2003vj,Kovtun:2012rj}. 

Turning to $\Sigma^{\rm aa}_{\rm AA}$, a similar treatment results in
\begin{align}
\label{Sigma-AA-0}
\Sigma_{\rm AA}^{\rma\rma}(\omega,\zev) = 2 g^{2} T^2 r^2(u_n^2-u_\gamma^2) \,{\rm Im} \int_{\vq} \frac{1}{\o+\i r+2\i D \qv^2}  \,.
\end{align}
Therefore, $\Delta \gamma(\omega)$, defined by eq.~(\ref{Delta-gamma}), reads
\begin{align}
\label{Delta-r-omega}
\Delta \gamma (\o)=-\frac{g^2 T^2 \chi r^2 (u_n^2-u_\gamma^2)} {8 \pi D} \,{\rm Re} \sqrt{\frac{r-\i \o }{2D}}\, .
\end{align}
We shall interpret our results for $\Delta \gamma (\o)$ at the end of this section. 

Finally, we turn to $\Sigma_{\rm VA}^{\rma\rmr}$ and $\Sigma_{\rm AA}^{\rma\rmr}$,
which can be obtained from the familiar steps. 
They are
\begin{align}
\label{tSigma-AA-0}
\Sigma_{\rm AA}^{\rma\rmr}(\omega,\zev) &= - g^{2} T^2 r (u _n-u _\gamma)  \int_{\vq} \frac{ r (u _n-u _\gamma)+ D \qv^2 (2 u _n)}{\o+\i r +2\i D \qv^2} \,, \\
\label{tSigma-VA-0}
\Sigma^{\rma\rmr}_{\rm VA}(\omega,\kv) &= -\i g^{2} T^2 u_n (\vv \cdot \kv)  \int_{\vq} \frac{ r (u_n-u_\gamma)+ D \qv^2 (2 u_n)}{\o+\i r +2\i D \qv^2} +\mathcal O(k^2)\,,
\end{align}
which, upon using eqs.~\eqref{Delta-r} and \eqref{Delta-def-end}, gives the expressions,
\begin{align}
\label{Delta-tr-omega}
\Delta r(\o)
&=- \frac{ g^2 T^2 r (u_n-u _\gamma) (r u _\gamma - \i \omega u_n )}{8 \pi D } {\rm Re}\sqrt{\frac{r-\i\o}{2D}}\,, \\
\label{Delta-alpha-omega}
\vv_{\rmA} (\o) &= -\frac{ g^2 T^2 \vv u_n (r u_\gamma -\i \omega u_n )}{8 \pi D } {\rm Re}\sqrt{\frac{r-\i\o}{2D}} \,.
\end{align}

We now have all the ingredients needed to compute the conductivity from eq.~(\ref{eq:Sigma-expr}). 
By the substitution of eqs.~(\ref{Delta-D-omega}), (\ref{Delta-r-omega}), (\ref{Delta-tr-omega}), and (\ref{Delta-alpha-omega}) into eq.~\eqref{eq:Sigma-expr}, we find $\s_{\perp}$ does not receive any loop corrections while $\sigma_{\parallel}$ at the one-loop order is given by
\begin{align}
\label{eq:Sigma-r-fin}
\sigma_{\parallel} (0)
& =(\s)_{0}+\frac{C^2 {\bm B}^2}{\chi r} \left[ 1 - \frac{g^2 T^2 \left( 2 u_n^2 + u_\gamma^2 \right)}{4\pi }\left (\frac{r}{2D}\right)^{\frac{3}{2}} \right] \,,
\end{align}
where the first and second terms in $\le[\ldots\ri]$ correspond to the CME related tree contribution (see eq.~(\ref{sigma-tree})) and the one-loop correction, respectively. Note that the fluctuation correction to $\sigma_{\parallel}$ is of the opposite sign to the tree one and lead to a positive-magnetoresistance contribution. Equation~\eqref{eq:Sigma-r-fin} is the main result of this section (see section~\ref{sec:conclusion} for further discussions).

Before closing this section, we remark that although the loop corrections to $\s$ are resulting from the combination of the set of functions listed in eq.~\eqref{RG-list}, our expression for each of them might be of interest on its own. Let us give two examples below.

For the first example, we recall that for a chiral medium microscopically described by non-Abelian gauge theories, $\g$ is referred to as Chern-Simons (CS) diffusion rate. In the weak-coupling regime, the CS diffusion rate is computed by accounting for contributions at the microscopic length scale $g^2T$ and time scale $g^4T$ \cite{Arnold:1996dy}.
Our result~\eqref{Delta-r-omega} can be interpreted as the additional contribution to the CS diffusive rate from the macroscopic length scale $\sim \sqrt{r/D}$. 

As for the second example, 
we consider the pole of $G^{\rm ra}_{\rm AA}$ in the limit $\kv \rightarrow \zev$, which, at the tree level, is located at $\omega=- \i r$. 
The one-loop corrections to this relaxation pole can be determined using the Dyson equation (\ref{eq:Dyson-Gra}),
\begin{align}
\label{eq:pole-GraAA}
(\mathcal G^{\rmr\rma}_{\rmA\rmA})^{-1}(\omega,\zev) =0\longrightarrow \omega + \i [r +\Delta r(\omega)] =0 \,.   
\end{align}
Upon substituting eq.~\eqref{Delta-tr-omega}, 
we find
\begin{align}
\omega = -\i r + \mathcal O(g^4)\,.
\end{align}
This should be contrasted with the naive expectation that the correction is of the order $g^{2}$.

\section{Conductivity at zero relaxation rate}
\label{sec:zeror}

To complement the analysis in the previous section, we shall work on the limit that the axial relaxation rate $\g$ vanishes in this section. By sending $r\to 0$ in the tree-level result \eqref{sigma-tree}, we notice that the CME-related contribution to the conductivity vanishes. However, as we shall show below, there is a CME-related contribution to the conductivity due to the effects of the fluctuations.

Instead of extracting conductivity tensor $\s^{ij}$ from relevant self-energies as was done in the previous section and in appendix~\ref{sec:alt-deri}, in this section, we will employ a complementary approach which determines $\s^{ij}$ from the retarded correlator of the vector current:
\begin{align}
\label{Kubo-R}
\sigma^{ij}(\omega)=\lim_{\vk\to \zev} \frac{1}{\omega}{\rm Im}\, {\CC}^{ij}_{\rm R}(K)\,.
\end{align}
Here, we have used the fluctuation-dissipation relation
\begin{align}
\label{FDT}
{\cal C}^{ij}(K)=   \frac{2T}{\o}{\rm Im}\,\CC^{ij}_{\rm R}(K)\,. 
\end{align}

Let us begin with the retarded correlator expressed in terms of the vector currents ${\bm J}^{\rmr}$ and ${\bm J}^{\rma}$:
\begin{align}
\label{CR-jra}
\CC^{ij}_{\rm R}(K)=\i \langle J^{i,\rmr} (K) J^{j,\rma}(-K)\rangle\,. 
\end{align}
Here and hereafter, $\langle\ldots\rangle$ denotes the average weighted by the path-integral \eqref{I-Z} with $\g=0$. At one-loop order, $\CC^{ij}_{\rm R}$ is given by the correlation of ${\bm J}^{\rm r}$ and ${\bm J}^{\rm a}$ expanded to the quadratic order in the fluctuating fields. In details, 
we use the expressions given by (\ref{jr}) and (\ref{ja}) with $\Ev=\zev$ and have
\begin{align}
\label{CR-loop}
\left[C^{ij}_{\rm R}(K)\right]_{{\rm loop}}&= {\rm i}\langle (J^{i,\rmr})_{1}(K)\, (J^{j,\rma})_{2}(-K)\rangle+ {\rm i}\langle (J^{i,\rmr})_{2}(K)\, (J^{j,\rma})_{1}(-K)\rangle
+ {\rm i}\langle (J^{i,\rmr})_{2}(K)\, (J^{j,\rma})_{2}(-K)\rangle\, ,
\end{align}
where
\bes
\label{jra-exp}
\begin{align}
({\bm J}^{\rmr})_{1}&=  C\vB (\mu_{\rm A})_{1}+{\cal O}(\bm \nabla) \,,
\\
({\bm J}^{\rmr})_{2}&= C (\mu_{\rm A})_{2}\vB- (\s_{\rm V})_{1}{\bm \nabla}(\mu_{\rm V})_{1}+2 \i T(\s_{\rm A})_{1} \na \psi_{\rm V} \,,
\\
({\bm J}^{\rma})_{1}&= C \pd_{t} \psi_{\rm A}\vB +{\cal O}(\bm \nabla)\,, 
\\
({\bm J}^{\rma})_{2}&=\pd_{t}\le[(\s_{\rm V})_{1}{\bm \nabla}\psi_{\rm V} \ri] +{\cal O}(\bm \nabla)\,.
\end{align}
\ees
Here, $\mathcal O(\bm \nabla)$ denotes terms which would vanish in the limit $\vk \rightarrow \zev$. 

To proceed further, we first show that the first two terms in eq.~\eqref{CR-loop} do not contribute to  eq.~\eqref{CR-jra} for $\vk \rightarrow \zev$:
\begin{align}
\lim_{\vk\to \zev}\langle (J^{i,\rmr})_{1}(K)  (J^{j,\rma})_{2}(-K)\rangle=0\,,
\quad
\lim_{\vk\to \zev}\langle (J^{i,\rmr})_{2}(K)  (J^{j,\rma})_{1}(-K)\rangle=0\,.
\end{align}
To see this, we consider
\begin{align}
\label{l-j2}
\langle\l_{\a}(x) ({\bm J}^{\rma})_{2}(x')\rangle
&=\sum_{\beta} \left( \int_{y}G^{\rm rr}_{\a\b}(x,y)\left< \frac{\d I_{{\rm int}}}{\d \l_{\b}(y)} ({\bm J}^{\rma})_{2}(x') \right>_0 
+ G^{\rm ra}_{\a\b}(x,y)\left< \frac{\d I_{{\rm int}}}{\d \psi_{\b}(y)} ({\bm J}^{\rma})_{2}(x') \right>_0 \right)
\no\\
&=\sum_{\beta} \int_{y}G^{\rm ra}_{\a\b}(x,y)\left< \frac{\d I_{{\rm int}}}{\d \psi_{\b}(y)} ({\bm J}^{\rma})_{2}(x') \right>_0 \,.
\end{align}
Here,  $I_{\rm int}=\int_{x}(\sL)_{3}$ and $\left< ... \right>_0$ denotes the average weighted by the Gaussian part of the effective action, as this suffices to the present one-loop calculations. 
In eq.~\eqref{l-j2}, we have also used 
\begin{align}
\left< \frac{\d I_{{\rm int}}}{\d \l_{\b}(y)} ({\bm J}^{\rma})_{2}(x')\right>_0 =0 \,,
\end{align}
which can be shown by causality (see ref.~\cite{Gao:2018bxz} for a general discussion) and the fact that both $\d I_{{\rm int}}/\d \psi_{\b}$ and $({\bm J}^{\rma})_{2}$ contain at least one power of the a-field.
On the other hand, in the absence of $\g$, one can confirm from eq.~\eqref{action} that $\d I_{{\rm int}}/\d \psi_{\b}=\na \cdot ({\bm J}^\rmr_\beta)_2$, 
meaning the Fourier transform of eq.~\eqref{l-j2} vanishes in the small $k$ limit.  
Therefore, the first term in eq.~\eqref{CR-loop}, which can be expressed as a linear combination of $\langle\l_{\a}(x)\, ({\bm J}^{\rm a})_{2}(x')\rangle $, should also vanish in this limit. 
A similar analysis applies to the second term in eq.~\eqref{CR-loop}. 
As a consequence, at one-loop order, eq.~\eqref{CR-loop} is reduced to
\begin{align}
\label{CR-loop0}
\lim_{\vk\to \zev}{\cal C}^{ij}_{\rm R}(K)= \i \lim_{\vk\to \zev} \left<(J^{i,\rmr})_{2}(K)  (J^{j,\rma})_{2}(-K) \right>_0 \,.
\end{align}

To compute eq.~\eqref{CR-loop0}, 
we rewrite $({\bm J}^{\rm r})_{2}$ and $({\bm J}^{\rm a})_{2}$ in terms of the rescaled fields (\ref{rescale}) using  eq.~(\ref{eq:couplings}):
\begin{align}
\label{j2-expr}
({\bm J}^{\rmr})_{2} &= 
- \frac{T\vv}{2}\, \tilde{u}_{n}\le(\l^{2}_{\rm V}+\l^{2}_{\rm A}\ri)
-T D_\rmV\tilde{u}_{\s} \lambda_\rmA \left( \na\lambda_{\rm V} - 2 \i \na \psi_\rmV \right) \,, 
\\
\label{j2-expa}
 ({\bm J}^{\rma})_{2} &= D_\rmV \tilde{u}_{\s} \pd_{t}\le(\lambda_\rmA \na \psi_{\rm V}\ri) \,, 
\end{align}
where we have further assumed the absence of the background vector chemical potential,
\begin{align}
\label{eq:back-A}
(n_\rmV)_0 =0\,, \quad (n_\rmA)_0\neq 0\,,
\end{align}
so that  $n_{\rm A;VA}=n_{\rm A;AV}=0$ and $\sigma_{\rmV;\rmV}=0$. Here, the parameters which describe the strength of non-linearity are defined by
\begin{align}
\label{eq:util}
\tilde{u}_{n}\equiv n_{\rm A;VV}=n_{\rm V;AA}\,,
\quad
\tilde{u}_{\s}\equiv \sigma_{\rmV;\rmA}\,. 
\end{align}

Noting the correlation between the last term in eq.~\eqref{j2-expr} and eq.~\eqref{j2-expa} vanishes by causality,
we now have two remaining contributions to eq.~\eqref{CR-loop0}: the correlations between $({\bm J}^a)_2$ and the first term (CME part) and the second term (diffusive part) of eq.~\eqref{j2-expr}. We shall first show that the former contribution vanishes for $\vk \rightarrow \zev$. 
Indeed,
\begin{align}
\label{VV cal}
& \lim_{\vk\to \zev}\langle (\l_{\rm V}\l_{\rm V})(K) (\l_{\rm A}\na\l_{\rm V})(-K)\rangle
\no \\
&=\int_{\vq}\sum_{m,m'}\i \qv \le[  R^{m}_{\rm VA}(\vq) R^{m'}_{\rm VV}(-\vq)
-
R^{m}_{\rm VV}(\vq) R^{m'}_{\rm VA}(-\vq)
\ri] \Delta^{mm'}(\o,\vq)
\,,
\end{align}
and similarly,
\begin{align}
\label{AA cal}
& \lim_{\vk\to \zev}\langle (\l_{\rm A}\l_{\rm A})(K) (\l_{\rm A}\na\l_{\rm V})(-K)\rangle
\no \\
&= \int_{\vq}\sum_{m,m'} \i \qv \le[ R^{m}_{\rm AA}(\vq) R^{m'}_{\rm AV}(-\vq)
-
R^{m}_{\rm AV}(\vq) R^{m'}_{\rm AA}(-\vq)
\ri] \Delta^{mm'}(\o,\vq)
\,.
\end{align}
By evaluating eq.~\eqref{R-pole} at $r=0$ and further assuming $D_{\rm V}=D_{\rm A}$,
we find
    \begin{align}
\label{R0}
R^{\pm}(\vk)=
\frac{1}{2}
\left(\begin{array}{cc}
1 & \pm {\rm sgn}(\vv\cdot\vk) \\
\pm {\rm sgn}(\vv\cdot\vk) & 1  
\end{array}
\right) \,.
\end{align}
Substituting eq.~\eqref{R0} into eqs.~\eqref{VV cal} and \eqref{AA cal} leads to
\begin{align}
& \lim_{\vk\to \zev}\langle (\l^{2}_{\rm V}+\l^{2}_{\rm A})(K)  (\l_{\rm A}\na \psi_{\rm V})(-K)\rangle= 0\,. 
\end{align}

The one-loop corrections to the conductivity tensor now becomes 
\begin{align}
\label{sigma-1}
\left[\s^{ij}(0)\right]_{{\rm loop}}&=T D^{2}\tilde{u}^{2}_{\s} \,{\rm Im} \lim_{\o\to 0}\lim_{\vk\to \zev}
\left< (\l_{\rm A}\pd^{i}\l_{\rm V})(K) (\l_{\rm A}\pd^{j}\psi_{\rm V})(-K) \right>
\no \\
&=
T D^{2}\tilde{u}^{2}_{\s}\,{\rm Im} \int_{Q}
q^{i}q^{j} \le[
G^{\rm rr}_{\rm AA}(Q) G^{\rm ra}_{\rm VV}(-Q)-
G^{\rm ra}_{\rm AV}(Q) G^{\rm rr}_{\rm VA}(-Q))
\ri]
\no \\
& =
T D^{2}\tilde{u}^{2}_{\s} \,{\rm Im}\, \i \int_{Q} q^{i}q^{j}  
\le[
G^{\rm ra}_{\rm AA}(Q) G^{\rm ra}_{\rm VV}(-Q)-
G^{\rm ra}_{\rm AV}(Q) G^{\rm ra}_{\rm VA}(-Q)
\ri]
\no \\
&\,
=T D^{2}\tilde{u}^{2}_{\s} \, {\rm Im}\, \int_{\vq}
\sum_{m,m'} q^{i}q^{j}
\le[
R^{m}_{\rm AA}(\vq) R^{m'}_{\rm VV}(-\vq)-
R^{m}_{\rm AV}(\vq) R^{m'}_{\rm VA}(-\vq)
\ri] \Delta^{mm'}(0,\vq)\,. 
\end{align}
It is straightforward to show from eq.~\eqref{R0} that the terms inside $[...]$ of eq.~\eqref{sigma-1} vanish when $(m,m')=(+,-), (-,+)$. On the other hand, the contributions from $(m,m')=(+,+), (-,-)$ give a finite result with 
\begin{align}
\Delta^{mm}(0,\qv)= \frac{1}{2\i D\qv^2 - 2m |{\vv\cdot\qv}| }\,.
\end{align}
Thus, we finally have
\begin{align}
\label{sigma-2}
\left[\s^{ij}(0)\right]_{{\rm loop}}&= - \frac{T(D\tilde{u}_{\sigma} )^2 }{D} \int_{\qv} \frac {D^2 \qv^2 q^i q^j }{2[(D\qv^2)^2 + (\vv\cdot\qv)^2 ]}
\no \\
&=-\frac{T (D\tilde{u}_{\sigma} )^2 }{192 \pi D}
\left(\frac{|\vv|}{D} \right)^{3}
\le[ 4 \hat v^i \hat v^j +(\delta^{ij}-\hat v^i \hat v^j)\ri]\,. 
\end{align}
From the first line to the second line in eq.~\eqref{sigma-2}, we have first used the dimensional regularization to integrate $q$ and then integrate over the angle $\hat{\Bv}\cdot \hat \vq$ analytically. 
The result of doing so gives rise to an emergent IR scale, 
\begin{align}
q_{*B} = \frac{|\vv|}{D}\,.
\end{align}
In our counting scheme ${\bm B} = {\cal O}(\epsilon)$, it follows that $|\vv|\ll 1$, and one can again verify that $q_{*B} \ll \Lambda$.

Finally, by comparing eq.~\eqref{sigma-2} with eq.~\eqref{eq:sigma-decomp},
we find the main results of this section,
\begin{align}
\label{sigma-zeror}
\frac{\Delta \sigma_{\rmV\parallel} (0)}{(\sigma)_{0}}=\frac{4\Delta \sigma_{\rmV\perp} (0)}{(\sigma)_{0}} = - \frac{(gT \tilde{u}_{\s})^2 }{48 \pi}\left(\frac{|\vv|}{D} \right)^{3} \,.
\end{align}
See the subsequent section for the discussion of eq.~\eqref{sigma-zeror}. 

In ref.~\cite{Delacretaz:2020jis}, the authors consider the fluctuation effects of a single chiral charge in the presence of the CME. 
For spatial dimension $d=3$, they find finite corrections to the AC conductivity ($\o\neq 0$) but vanishing DC conductivity ($\omega=0$). 
The differences between theirs and the present results mainly arise from the fact that we have considered the couplings between the axial and vector charge densities (see also refs.~\cite{Kovtun:2014nsa,Chen-Lin:2018kfl,Mukerjee_2006} on the studies of fluctuation dynamics with multiple conserved charges).

\section{Discussion}
\label{sec:conclusion}

\subsection{Positive magnetoresistance}

We presented in this paper the diagrammatic calculation of the modifications of conductivity (the inverse of resistance) for a chiral medium with the chiral magnetic effect (CME) based on the non-equilibrium  effective field theory (EFT) approach. 
We consider a generic vector charge density and axial charge density as slow variables and study the intertwined effects from their fluctuations and CME. 
For the first time, we obtain the CME-related modifications to the conductivity tensors due to fluctuations for systems with finite and vanishing axial relaxation rate $r$, as summarized in eq.~\eqref{eq:Sigma-r-fin} and eq.~\eqref{sigma-zeror}, respectively. 

Contrary to the common statement that the CME leads to a negative magnetoresistance, we find that CME-related effects due to fluctuations give rise to a {\it positive} magnetoresistance. 
Whether the {\it net} magnetoresistance is positive or negative is determined by the competition between the two terms inside $[...]$ in eq.~\eqref{eq:Sigma-r-fin}.
Note, to apply our one-loop results, we should require the fluctuation contribution to be much smaller than the tree-level expression, which includes both the first term and second term on the right-hand side of eq.~\eqref{eq:Sigma-r-fin} (c.f.~eq.~\eqref{sigma-tree}). 
Since we are working in the weak $\Bv$ limit, the tree-level contribution is dominated by $(\sigma)_0$, which is indeed much larger than the one-loop correction, i.e., the third term on the right-hand side of eq.~\eqref{eq:Sigma-r-fin}.
However, this does not necessarily mean that the third term has to be smaller than the second term. Therefore, the net CME-related contribution can in principle be dominated by the fluctuations effects. We have made a number of simplifications in our analysis, but we hope this qualitative feature might have some implication to real physical systems.
Interestingly, a positive magnetoresistance might have already been seen in Weyl semimetals when the magnetic field is small (e.g., ref.~\cite{Li:2014bha}).

It is truly striking that the CME contributes to the conductivity even in the limit $r=0$ (see eq.~\eqref{sigma-zeror}). 
This is in a marked difference from the result, which does not account for fluctuations that the CME contribution vanishes in this limit. 
Moreover, the fluctuation modifies both longitudinal and transverse conductivities, while at finite $r$, only longitudinal conductivity receives the correction from the CME.

The parametric behavior of the ratio of fluctuation effects to the tree-level contribution is very instructive. 
From eqs.~\eqref{eq:Sigma-r-fin} and \eqref{sigma-zeror}, we schematically have
\begin{align}
\label{eq:ratio}
\frac{\rm (Hydrodynamic\ fluctuation)}{\rm (Bare \ contribution)} \sim  (T u)^2  g^2 q_{*}^3\,,
\end{align}
where $q_{*}$ is characteristic momentum of fluctuating modes. For the case with $r \neq 0$,  $q_{*}=\sqrt{r/D}$, resulting from the competition between the diffusion of vector charge and the damping of the axial charge, whereas for $r=0$, $q_{*}=|{\bm v}|/D$ originating from the competition between diffusion and propagation of the chiral magnetic wave (CMW). 
Equation \eqref{eq:ratio} indicates that the relative importance of fluctuation corrections is determined by two factors, a) the strength of non-linearity $(T u)^2$ and b) the ratio between the phase space volume of the long-wavelength fluctuating modes, $q^3_{*}$, and that of the whole system, $g^{-2}=\chi T$.

Let us end this section by comparing the role the CME played in two cases under study. 
For $r \neq 0$, the CME gives rise to non-linear couplings among charge fluctuations but plays no role in determining the IR scale. 
In contrast, for $r = 0$ and at vanishing background vector charge, the non-linearity relevant to the finite corrections to the conductivity solely comes from the density-dependence of diffusive constant and conductivity but does not rely on the CME. 
However, the competition between CMW propagation and charge diffusion leads to the emergent IR scale. 
Due to the differences explained above, $C|\Bv|$ dependence of $\Delta\s$ is different. The correction to the conductivity scales as $(C\Bv)^2$ in the former case and scales as $(C|\Bv|)^{3}$ in the latter case.

\subsection{Remarks on hydrodynamic fluctuations}

Another motivation of the present study is to advance our understanding of general aspects of hydrodynamic fluctuations. 
Here, we employ the recently developed non-equilibrium EFT for the present studies. 
Our exercise here demonstrates this EFT approach allows us to use powerful (and familiar) field theory techniques to analyze hydrodynamic fluctuations. By construction, the EFT automatically takes into account the constraints from symmetries. 
For example, in appendix~\ref{sec:alt-deri}, we show explicitly how fluctuations-dissipation theorem and the Ward-Takahashi identity are satisfied at one-loop order. 
In the traditional method, special care is needed to ensure those relations~(see the recent work \cite{Chao:2020kcf} for the former). 

Remarkably, the fluctuation contribution to the conductivity is finite due to the CME. This should be contrasted with the case of an ordinary fluid that fluctuation corrections to transport coefficients (at zero frequency limit) are typically zero.  
Such difference is related to the emergent IR momentum scale behavior in the loop integration, $q_{*}$. 
Generically, the corrections to transport coefficients should scale with $q_{*}$ to an appropriate (and positive) power for $d=3$. 
For a normal fluid, $q_{*}\sim \sqrt{\o}$, giving rise to the renowned long-time tail phenomena~\cite{POMEAU197563,Kovtun:2003vj,Kovtun:2012rj}, and consequently vanishes in the limit $\omega \rightarrow 0$ (see also refs~\cite{Akamatsu:2016llw,Stephanov:2017ghc,Jain:2020hcu} for related discussion). 
Therefore, to obtain finite corrections to transport coefficients in this limit, there must be additional soft scales. 
Such scales are generated by the magnetic field and/or axial relaxation rate in the present study. 
Given the generality of the discussion above, we anticipate that the CME and hydrodynamic fluctuations together might contribute to other transports coefficients. 
A natural follow-up would be to include fluctuations from energy and momentum densities and study the effects on shear and bulk viscosities. 
We leave these and other extensions of this work to future studies.

\acknowledgments
We thank Kenji Fukushima, Yoshimasa Hidaka, Dmitri Kharzeev, Chris Lau and Derek Teaney for useful discussions and comments.
N.~S. and Y.~Y. would like to acknowledge financial support by the Strategic Priority Research Program of Chinese Academy of Sciences, Grant No. XDB34000000.
N.~Y. was supported by the Keio Institute of Pure and Applied Sciences (KiPAS) project at Keio University and JSPS KAKENHI Grant No.~19K03852.

\appendix

\section{Derivation of $I_{{\rm anom}}$}
\label{sec:Anomaly-matching}

In this section, we shall present the derivation of $I_{{\rm anom}}$ in eq.~\eqref{L-anom} used in the main text following the method of ref.~\cite{Glorioso:2017lcn}. 
We shall see the anomaly relation combined with KMS invariance uniquely fix the form of CME. 
This should be contrasted with the derivation of CME in hydrodynamics from the second law of thermodynamics~\cite{Son:2009tf,Hattori:2017usa}. 

Let us first consider the low-energy effective action describing the slow mode associated with a single chiral charge. 
We consider the action divided into two parts,
\begin{align}
    I=I_{{\rm inv}}+I_{{\rm anom}}\,,
\end{align}
where $I_{{\rm inv}}$ is identical to the hydrodynamic action of a conserved charge (see eq.~\eqref{L-norm-V}). We shall focus on the anomaly-related action $I_{{\rm anom}}$ from now on. 

Because of the anomaly, 
the consistent $\rmr$-current, 
\begin{align}
J^{\mu}=\frac{\d I}{\d A^{\rm a}_{\mu}}\, 
\end{align}
obeys the anomaly equation 
\begin{align}
\label{eq:consistent-anomaly-k}
\partial_\mu J^{\mu} = -\frac{\kappa C}{24} F_{\mu \nu} \tilde F^{{\mu \nu}}\,,
\end{align}
where  $\kappa=\pm1$ correspond to the right-handed and left-handed charge, respectively. 
In this appendix, we shall keep the index ``$\rma$'' but suppress the index ``$\rmr$.'' 
To obtain eq.~\eqref{eq:consistent-anomaly-k}, 
we use the consistent anomaly equation on the two segments of Schwinger-Keldysh contour labeled by ``1,2'':
\begin{align}
\partial_\mu J^{\mu}_{1} = -\frac{\kappa C}{24} F_{\mu \nu, 1} \tilde F_{1}^{\mu \nu}\, , 
\quad
\partial_\mu J^{\mu}_{2} = -\frac{\kappa C}{24} F_{\mu \nu, 2} \tilde F_{2}^{\mu \nu}\, , 
\end{align}
to compute $\pd_{\mu}J^{\mu}=\pd_{\mu}\le( J^{\mu}_{1}+J^{\mu}_{2}\ri)$. 
We have not included a contribution quadratic in $\rm a$-field on the right-hand side of eq.~\eqref{eq:consistent-anomaly-k} since such contribution should arise from the action involving three powers of $\rm a$-field.  
Similar to the discussion presented in the main text, the equation of motion for $\psi^{\rm a}$,
\begin{align}
    \frac{\d I}{\d \psi^{\rm a}}=0\,,
\end{align}
should be equivalent to the consistent anomaly equation \eqref{eq:consistent-anomaly-k}. 
Therefore, the anomaly part of the action takes the form 
\begin{align}
\label{eq:S-anom-k}
I_{{\rm anom}} &= 
\int_x \left(-\frac{ \kappa C}{24}  
\psi^\rma F_{\mu \nu} \tilde F^{\mu \nu} +  J_{\rm cons}^\mu \mathcal A ^\rma_\mu  \right)\,.
\end{align}
Here, ${\cal A}^{\rm a}_{\mu} = A^{\rm a}_{\mu}+\pd_{\mu}\psi^{\rm a}$ as given in eq.~\eqref{C-def} and $J_{\rm cons}$ ($J_{{\rm cov}}$) denotes the contribution to the consistent (covariant) current only from $I_{{\rm anom}}$, distinguished with the current $J$ from the total action $I$. The difference between $J_{\rm cons}$ and $J_{{\rm cov}}$ defines the Chern-Simons (CS) current as
\begin{align}
\label{j-cov}
J_{{\rm cov}}=J_{{\rm cons}}+J_{{\rm CS}}\, , 
\end{align}
where
\begin{align}
\label{j-CS}
J^{\mu}_{{\rm CS}}= -\frac{\kappa C}{6}\tilde{F}^{\mu\nu}\, A_{\nu}\, .
\end{align}
Similar to our previous treatment of the consistent anomaly equation \eqref{eq:consistent-anomaly-k}, 
we have dropped terms quadratic in a-fields in eq.~\eqref{j-CS}. 
Substituting eqs.~\eqref{j-CS} and \eqref{j-cov} into eq.~\eqref{eq:S-anom-k}, 
we now have
\begin{align}
\label{I-anom-2}
    I_{{\rm anom}}= \int_{x}\, \le(
    -\frac{C\kappa}{24}\psi^{\rm a} F_{\mu \nu} \tilde{F}^{\mu \nu} + \frac{\kappa C}{6} \tilde{F}^{\mu\nu}A_{\nu}{\cal A}^{\rm a}_{\mu}+ J_{{\rm cov}}^{\mu} {\cal A}^{\rm a}_{\mu}
    \ri)\, . 
\end{align}

 We shall consider the following form for $J^{\mu}_{\rm cov}$:
\begin{align}
J^\mu_{\rm cov} = \xi(\mu) B^{\mu}\,,
\end{align}
where $\xi(\mu)$ is a function of $\mu$ and we have defined $E^{\mu}=F^{\mu\nu} \ell_{\nu}$ and $B^{\mu}=\tilde{F}^{\mu\nu} \ell_{\nu}$ with $\ell^{\mu} = (1, {\bm 0})$ denoting the frame of the medium. 
We shall also use below that $\mu = \ell \cdot {\cal A}$ (see eq.~\eqref{mu-def}). 
Since $I_{{\rm anom}}$ is invariant under the KMS transformations~\eqref{KMS-r} and \eqref{KMS-a},
\begin{align}
{\cal A}^{\rm r}_{\mu}&\to\Theta {\cal A}^{\rm r}_{\mu}\, ,
    \quad
    \psi^{\rm r}\to \Theta \psi^{\rm r} \,,
  \no \\
  {\cal A}^{\rm a}_{\mu}&\to \Theta {\cal A}^{\rm a}_{\mu}+ \frac{\rm i}{T} \Theta \le(\ell^{\a}\pd_{\a}{\cal A}_{\mu}\ri)
  \no \\ 
  &=\Theta {\cal A}^{\rm a}_{\mu}+ \frac{\rm i}{T} \Theta \ell^{\a}\le(F_{\a\mu}+\pd_{\mu}{\cal A}_{\a}\ri)
    =\Theta {\cal A}^{\rm a}_{\mu}+ \frac{\rm i}{T} \Theta \le(-E_{\mu}+\pd_{\mu}\mu\ri)\, , \no
    \\
   \psi^{\rm a}&\to \Theta \psi^{\rm a}+ \frac{\rm i}{T} \Theta \le(
    \ell^{\a}\pd_{\a}\psi
    \ri)\, . 
\end{align}
The variance of the first two terms in eq.~\eqref{I-anom-2} under the KMS transformation should precisely cancel that from the last term. 
This requirement uniquely fixes $\xi(\mu)$, as we shall explain below. 

Indeed, for $\Theta={\cal CPT}$, we have
\begin{align}
\d_{{\rm KMS}}\, \int_{x}\, \psi^{\rm a}F_{\mu \nu}\tilde{F}^{\mu \nu}= - \frac{\rm i}{T}\, \int_{x}\,(\ell^{\a}\pd_{\a}\psi) F_{\mu \nu}\tilde{F}^{\mu \nu}\, , 
\end{align}
where we have used $\Theta \mathcal A^\mu (x) = -\mathcal A^\mu (-x)$, $\Theta \tilde F ^{\mu\nu} (x) = \tilde F ^{\mu\nu}(-x)$, and $\Theta \ell^\mu (x) = \ell^\mu (-x)$. 
Similarly,
\begin{align}
&    \,\d_{{\rm KMS}}\int_{x}\, \tilde{F}^{\mu\nu}A_{\nu} {\cal A}^{\rm a}_{\mu}
= - \frac{\rm i}{T} \int_{x} \tilde{F}^{\mu\nu}A_{\nu} (-E_{\mu}+\pd_{\mu}\mu)
= \frac{\rm i}{T} \int_{x} 
\le[\frac{1}{4}(\ell \cdot A)+\frac{\mu}{2}\ri] F_{\mu \nu} \tilde{F}^{\mu \nu}\,,
\end{align}
where we have used the identity
\begin{align}
\tilde{F}^{\mu\nu}A_{\nu} E_{\mu}
=\le(\ell^{\mu}B^{\nu}-\ell^{\nu}B^{\mu}+\epsilon^{\mu\nu\a\b}\ell_{\a}E_{\b}\ri)\,A_{\nu} E_{\mu}
=- B\cdot E (\ell \cdot A) = \frac{1}{4}F_{\mu \nu} \tilde{F}^{\mu \nu} (\ell \cdot A)\, . 
\end{align}
Putting all pieces together, we find 
\begin{align}
\label{KMS-A1}
&\,     \delta_{{\rm KMS}} \int_{x}\, \le(
-\frac{C\kappa}{24}\psi^{\rm a} F_{\mu \nu} \tilde{F}^{\mu \nu} + \frac{\kappa C}{6} \tilde{F}^{\mu\nu}A_{\nu}{\cal A}^{\rm a}_{\mu} \ri)
= \frac{{\rm i} \kappa C}{6T}\int_{x}\le[
\frac{1}{4} (\ell \cdot \pd)\psi^{\rm a} +\frac{1}{4}(\ell \cdot A)+\frac{\mu}{2}
\ri]F_{\mu \nu} \tilde{F}^{\mu \nu}
\no \\
&= \frac{{\rm i} \kappa C}{8T}\int_{x}\mu F_{\mu \nu} \tilde{F}^{\mu \nu}\, . 
\end{align}
On the other hand,
\begin{align}
\label{KMS-A2}
\delta_{{\rm KMS}}\, \int_{x}J^\mu_{\rm cov} {\cal A}^{\rm a}_{\mu}
= \frac{\rm i}{T} \int_{x}\xi(-\mu)B^{\mu}\le(-E_{\mu}+\pd_{\mu}\mu\ri)
=\frac{\rm i}{4T}\int_{x}\xi(-\mu) F_{\mu \nu} \tilde{F}^{\mu \nu}\,, 
\end{align}
where we used the fact that $\xi(-\mu)B^{\mu}\pd_{\mu}\mu$ is a total derivative.
For eq.~\eqref{KMS-A1} to cancel eq.~\eqref{KMS-A2}, we have the appropriate form for the CME:
\begin{align}
\label{eq:jcon-expr}
\xi(\mu)= \frac{\kappa C}{2}\mu\,, \quad J^{\mu}_{\rm cov}= \frac{\kappa C}{2}\mu B^{\mu}\,. 
\end{align}

Now, we generalize the discussion above to the system with both axial and vector charges. 
From the consistent anomaly relation,
\begin{align}
\partial_\mu J^{\mu}_{\rmV} &=0\,, \\
\partial_\mu J^{\mu}_{\rmA} &= -\frac{C}{12} \left( 3F_{\mu\nu,\rmV} \tilde F^{\mu\nu}_\rmV + F_{\mu\nu,\rmA} \tilde F^{\mu\nu}_\rmA \right)\, .
\end{align}
Thus, we have, in analogous to eq.~\eqref{eq:S-anom-k}, the expression for $I_{{\rm anom}}$:
\begin{align}
\label{I-anom-full}
I_{\rm anom} &=
\int_x \left[-\frac{C}{12}
\psi_\rmA^\rma \left( 3F_{\mu\nu,\rmV} \tilde F^{\mu\nu}_\rmV + F_{\mu\nu,\rmA} \tilde F^{\mu\nu}_\rmA \right) \right. \notag\\
& \qquad \quad \left. + \left(J^\mu_{\rm cov,\rmV} + C \tilde F^{\mu\nu}_\rmV A_{\nu, \rmA} \right) \mathcal A^\rma_{\mu,\rmV} + \left(J^\mu_{\rm cov,\rmA} + \frac{C}{3}\tilde F^{\mu\nu}_\rmA A_{\nu, \rmA} \right) \mathcal A^\rma_{\mu,\rmA}  \right] \,.
\end{align}
By imposing the KMS symmetry to eq.~\eqref{I-anom-full}, we find
\begin{align}
\label{eq:j-cov-V-A}
J^\mu_{\rm cov,V} = C  \left( \mu_{\rmA} B^{\mu}_\rmV 
+ \mu_{\rmV} B^{\mu}_\rmA 
\right) \,,\quad
J^\mu_{\rm cov,A} = C \left( \mu_{\rmV} B^{\mu}_\rmV 
+ \mu_{\rmA} B^{\mu}_\rmA 
\right)\,.
\end{align}
Substituting eq.~\eqref{eq:j-cov-V-A} into eq.~\eqref{I-anom-full} leads to the desired expression for $I_{{\rm anom}}$, which reduces to eq.~\eqref{L-anom} when the external axial gauge field is absent.

\section{One-loop expression for self-energy $\Sigma^{\rm 
aa}_{\a\b}$}
\label{sec:self-aa}

We here show the expression for $\Sigma^{\rm aa}_{\a\b}$ defined in eq.~\eqref{eq:Dyson-Grr} at the one-loop level.
There are three contributions
\begin{align}
\label{sigma-aa-all}
(\Sigma^{\rm aa}_{\a\b})=(\Sigma^{\rm aa}_{\a\b})_{\rm I}+(\Sigma^{\rm aa}_{\a\b})_{\rm II}+(\Sigma^{\rm aa}_{\a\b})_{\rm III}\, , 
\end{align}
which correspond to the diagrams given by figures~\ref{fig:Sigmas} (d), (e), and (f), respectively. 
Let us start with the first two contributions in eq.~\eqref{sigma-aa-all}:
\begin{align}
\label{eq:sigma-aa-I}
(\Sigma^{\rm aa}_{\alpha \beta})_{\rm I} (K) &= 4 T^2 \sum_{\gamma\delta\sigma\tau} \int_Q g_\alpha U^{-\kv;-\qv_-,\qv_+}_{\alpha;\gamma\sigma} G^{\rm rr}_{\sigma\tau} (q^0,\qv_+) G^{\rm ra}_{\gamma\delta} (\omega-q^0,-\qv_-) g_\delta V^{\qv_-,\kv;-\qv_+}_{\delta\beta;\tau}\,, \\
(\Sigma^{\rm aa}_{\alpha \beta})_{\rm II} (K) &= 4 T^2 \sum_{\gamma\delta\sigma\tau} \int_Q g_\gamma V^{-\kv,-\qv_-;\qv_+}_{\alpha\gamma;\sigma} G^{\rm rr}_{\sigma\tau} (q^0,\qv_+) G^{\rm ar}_{\gamma\delta} (\omega-q^0,-\qv_-) g_\beta U^{\kv;\qv_-,-\qv_+}_{\beta;\delta \tau}\,.
\end{align}
One can show that
\begin{align}
\label{eq:Sigma-aa-I-II}
(\Sigma^{\rm aa}_{\alpha \beta})_{\rm II} = (\Sigma^{\rm aa}_{\beta \alpha})_{\rm I}^* \,,
\end{align}
using eq.~(\ref{eq:Gra-Gar}),
\begin{align}
G^{\rm rr}(K) = (G^{\rm rr})^\dagger (K) = G^{\rm rr}(-K)\,,
\end{align}
and 
\begin{align}
\label{eq:UV-conj}
\left(U^{\kv_1;\kv_2,\kv_3}_{\alpha;\beta\gamma}\right)^* = - U^{-\kv_1;-\kv_2,-\kv_3}_{\alpha;\beta\gamma}  \,,\quad \left( V_{\rmA \alpha; \beta}^{\kv_1,\kv_2;\kv_3}  \right)^* = V_{\rmA \alpha; \beta}^{\kv_1,\kv_2;\kv_3} \,,
\end{align}
which can be verified from eq.~(\ref{UVs}).

To calculate $(\Sigma^{\rm aa}_{\alpha \beta})_{\rm I}$, we use eq.~(\ref{eq:FDT2}) and write $G^{\rm rr}$ as a sum of $G^{\rm ra}$ and $G^{\rm ar}$.
Then, we drop the latter contribution using eq.~(\ref{dq0-zero}) and carry out the $q^0$ integral from the former contribution using eq.~(\ref{q0-int}). We obtain 
\begin{align}
\label{eq:Sigma-aa-I-2}
(\Sigma^{\rm aa}_{\alpha \beta})_{\rm I} (K) &= -\i T^2 \sum_{\gamma\delta\sigma\tau} \sum_{mm'} \int_Q 
\left(F^{{\rm aa},mm'}_{\alpha\beta}\right)_{\rm I}(\kv,\qv)  \Delta^{mm'}(k,\qv)\,,
\end{align}
with
\begin{align}
\label{eq:F-aa-I}
\left(F^{{\rm aa},mm'}_{\alpha\beta}\right)_{\rm I} (\kv,\qv) = {\rm tr}\left[
R^{m} (\qv_+)
\left( g_\alpha U^{-\kv;-\qv_-,\qv_+}_{\alpha} \right)^{\rm T} R^{m'} (-\qv_-) \left( 4\i g V^{\qv_-,\kv;-\qv_+}_{\beta}\right) \right]\,,
\end{align}
where we have used the same matrix form as eq.~(\ref{Gamma-1}):
\begin{align}
\left( g_\alpha U^{-\kv;-\qv_-,\qv_+}_{\alpha} \right)_{\beta\gamma} = g_\alpha U^{-\kv;-\qv_-,\qv_+}_{\alpha;\beta\gamma} \,,\quad \left( 4\i g V^{\qv_-,\kv;-\qv_+}_{\beta}\right)_{\gamma\delta} = 4\i g_\gamma V^{\qv_-,\kv;-\qv_+}_{\gamma\beta;\delta} \,.
\end{align}
By using eq.~(\ref{eq:Sigma-aa-I-II}) we have
\begin{align}
\label{eq:Sigma-aa-I-II-expr}
(\Sigma^{\rm aa}_{\alpha \beta})_{\rm I}(K) + (\Sigma^{\rm aa}_{\alpha \beta})_{\rm II} (K) &= -\i  T^2 \sum_{mm'} \int_{\qv} \left[
\left(F^{{\rm aa},mm'}_{\alpha\beta}\right)_{\rm I} (\kv,\qv) \Delta^{mm'}(k,\qv) - (\alpha\leftrightarrow\beta+{\rm c.c.} ) \right]\,.
\end{align}
Here, $(\alpha\leftrightarrow\beta+{\rm c.c.} )$ denotes the complex conjugate of the first term with the interchanged ($\alpha,\beta$) labels. 

The third contribution, figure~\ref{fig:Sigmas} (f), can be written as  
\begin{align}
\label{eq:Sigma-aa-III}
(\Sigma^{\rm aa}_{\alpha\beta})_{\rm III} (K) &= 2 T^2 \sum_{\gamma\delta\sigma\tau} \int_Q g_\alpha U^{-\kv;-\qv_-,\qv_+}_{\alpha;\gamma\sigma} G^{\rm rr}_{\sigma\tau} (q^0,\qv_+) G^{\rm rr}_{\gamma\delta} (\omega-q^0,-\qv_-) g_\beta U^{\kv;\qv_-,-\qv_+}_{\beta;\delta\tau}\,.
\end{align}
Using eqs.~(\ref{eq:Gra-Gar}) and (\ref{eq:FDT2}), the $q^0$ integral relevant to (\ref{eq:Sigma-aa-III}) can be written as
\begin{align}
\int_{q^0} G^{\rm rr}_{\sigma\tau} (q^0,\qv_+) G^{\rm rr}_{\gamma\delta} (\omega-q^0,-\qv_-)
&= \i \int_{q^0} \left[\i G^{\rm ra}_{\sigma\tau} (q^0,\qv_+) G^{\rm ra}_{\gamma\delta} (\omega-q^0,-\qv_-) - ({\rm c.c.}) \right] \,.
\end{align}
Further using
\begin{align}
g_\alpha U^{-\kv;-\qv_-,\qv_+}_{\alpha;\gamma\sigma} g_\beta U^{\kv;\qv_-,-\qv_+}_{\beta;\delta\tau} &= \left(g_\beta U^{-\kv;-\qv_-,\qv_+}_{\beta;\delta\tau} g_\alpha U^{\kv;\qv_-,-\qv_+}_{\alpha;\gamma\sigma} \right)^*\,,
\end{align}
which can be readily checked from eq.~(\ref{eq:UV-conj}), we express eq.~(\ref{eq:Sigma-aa-III}) as
\begin{align}
\label{eq:Sigma-aa-III-expr}
(\Sigma^{\rm aa}_{\alpha\beta})_{\rm III} (K) &= 2\i T^2 \sum_{\gamma\delta\sigma\tau} \int_Q \left[ g_\alpha U^{-\kv;-\qv_-,\qv_+}_{\alpha;\gamma\sigma} G^{\rm ra}_{\sigma\tau}(q^0,\qv_+) \i G^{\rm ra}_{\gamma\delta} (\omega-q^0,-\qv_-)  g_\beta U^{\kv;\qv_-,-\qv_+}_{\beta;\delta\tau} \right. \notag\\
& \hspace{250pt} \left.
- (\alpha\leftrightarrow\beta+{\rm c.c.}) \right]\,.
\end{align}
Finally, we can combine eqs.~(\ref{eq:Sigma-aa-I-II-expr}) and (\ref{eq:Sigma-aa-III-expr}) into the form:
\begin{align}
\label{Sigma-aa-final-general}
(\Sigma^{\rm aa}_{\alpha\beta}) (K) &= -\frac{\i T^2}{2} \sum_{mm'} \int_{\qv} \left[
F^{{\rm aa},mm'}_{\alpha\beta}(\kv,\qv) \Delta^{mm'}(k,\qv) - (\,\alpha\leftrightarrow\beta\,+\,{\rm c.c.})  \right]\,,
\end{align}
where 
\begin{align}
F^{{\rm aa},mm'}_{\alpha\beta} (\kv,\qv) = {\rm tr}\left[
R^{m} (\qv_+)
\left( \Gamma^{\rma}_{\alpha} \right)^{\rm T} R^{m'} (-\qv_-) \Gamma'^\rma_\beta \right]\,,
\end{align}
with
\begin{align}
(\Gamma^\rma_\alpha)_{\beta\gamma} & = g_\alpha U^{-\kv;-\qv_-,\qv_+}_{\alpha;\beta\gamma} \,, \\
(\Gamma'^\rma_\beta)_{\gamma\delta} & = 4 \left(- g_\beta U^{\kv;\qv_-,-\qv_+}_{\gamma;\beta\delta} + 2 \i g_\gamma V^{\qv_-,\kv;-\qv_+}_{\gamma\beta;\delta} \right) \,.
\end{align}
If $\alpha=\beta$, eq.~(\ref{Sigma-aa-final-general}) reduces to the form (\ref{Sigma-aa-final}) in the main text.

\section{Conductivity tensor from density-density correlator at zero axial relaxation rate}
\label{sec:FDT-check}

We here show another derivation of the one-loop conductivity, using the symmetrized density-density correlator $\mathcal C^{00}_{\rm S} $, at the vanishing axial relaxation rate with zero background vector charge density but finite axial charge density. The derivation here is more parallel to the analysis in section \ref{sec:finite-r}, whereas it will give consistent results with those in section \ref{sec:zeror}, namely eq.~(\ref{sigma-zeror}).

First of all, by taking the limit $r\to 0$ followed by $\omega\to 0$, eqs.~\eqref{eq:Sigma-iso-expr} and \eqref{eq:Sigma-an-expr} are reduced to
\begin{align}
\s_{\perp}(0)= (\s_{\rm V})_{0}+\Delta \s_{\perp}(0)\,, 
\quad
\s_{\parallel}(0)= (\s_{\rm V})_{0}+\Delta \s_{\parallel}(0)\, , 
\end{align}
where $\Delta \s_{\perp}(0)$ and $\Delta \s_{\parallel}(0)$ can be obtained from eq.~\eqref{Delta-def-Sigma-aa-VV} by calculating  $\Sigma^{\rm aa}_{\rm VV}$ using eq.~(\ref{Sigma-aa-final}). 
In parallel to the analysis on eq.~(\ref{Gamma-small-k}), we consider $\Gamma^\rma_\rmV$ and $\Gamma^{\rma'}_\rmV$ defined in eqs.~(\ref{Gamma-1}) and (\ref{Gamma-2}) at the small $k$ limit:
\begin{align}
\label{eq:Gamma-1-r=0}
\Gamma^\rma_\rmV &= -\frac{\i g D}{2} (\kv\cdot\qv) \left(
\begin{array}{cc}
 0 & -\tilde u_\sigma \\
 \tilde u_\sigma & 0 
\end{array}
\right) - \frac{g}{2} (\vv_\rmA  \cdot \kv) \left(
\begin{array}{cc}
 \tilde u_n & 0\\
 0 & \tilde u_n 
\end{array}
\right) \,,\\
\label{eq:Gamma-2-r=0}
\Gamma^{\rma'}_\rmV &= 2\i g D  (\kv \cdot \qv) \left(
\begin{array}{cc}
 0 & 3 \tilde u_\sigma \\
 \tilde u_\sigma  & 0 
\end{array}
\right) - 2 g (\vv_\rmA\cdot \kv) \left(
\begin{array}{cc}
 \tilde u_n  & 0\\
 0 & \tilde u_n
\end{array}
\right)\, ,
\end{align}
where we have assumed eq.~\eqref{eq:back-A} and used eq.~\eqref{eq:util}.
It is easy to check that finite contribution to the DC conductivity comes from  $(m,m')=(+,+),(-,-)$ contribution with
\begin{align}
F^{\rm aa,++}_{\rmV \rmV} = F^{\rm aa,--}_{\rmV \rmV} =-(D g \tilde u_\sigma \kv\cdot\qv)^2 \, . 
\end{align}
We thus find  
\begin{align}
\label{eq:appC-results}
\Sigma^{\rm aa}_{\rm VV}(0,\kv) = (D g T \tilde u_\sigma)^2 \int_\qv \frac{ (\kv\cdot\qv)^2 D\qv^2 }{(D\qv^2)^2 + (\vv\cdot\qv)^2} + {\cal O}(k^3)\,.
\end{align}
Using eq.~(\ref{Delta-def-Sigma-aa-VV}), we obtain eq.~(\ref{sigma-2}) and resulting eq.~(\ref{sigma-zeror}), as it should be.

\section{Explicit verification of the fluctuation-dissipation relation and Ward-Takahashi identity at one-loop}
\label{sec:alt-deri}

In this appendix, we show explicitly that at one-loop order, the symmetrized current-current correlator ${\mathcal C}^{ij}$ is related to the retarded correlator ${\mathcal C}^{ij}_{\rm R}$ through the fluctuation-dissipation relation~\eqref{FDT}. 
Since we have already demonstrated that conductivity tensor obtained from ${\mathcal C}^{00}$ coincides with that from ${\mathcal C}^{ij}_{\rm R}$ in appendix~\ref{sec:FDT-check}, we therefore verify the constraint imposed by Ward-Takahashi identity at one-loop order between ${\cal C}^{00}$ and ${\cal C}^{ij}$. 

For illustrative purpose, we shall focus on the diffusive part of the current at quadratic order in fluctuations,
\begin{align}
\label{jr-diff}
({\bm J}^{\rmr})^{\rm dif}_2&=-T D_\rmV\tilde{u}_{\s} \lambda_\rmA \left( \na\lambda_{\rm V} - 2 \i \na \psi_\rmV \right) \,, \\
({\bm J}^{\rma})^{\rm dif}_2&=D_\rmV \tilde{u}_{\s} \pd_{t}\le(\lambda_\rmA \na \psi_{\rm V}\ri)\,. 
\end{align}
The second term in eq.~\eqref{jr-diff}, which is proportional to the a-field $\psi$, arises from the multiplicative noise.
We shall see this multiplicative noise contribution is crucial to ensure the fluctuation-dissipation relation and the Ward-Takahashi identity at one-loop order. 

We begin by computing the one-loop corrections to the symmetrized correlator
\begin{align}
\label{eq:4-sym-cor-dif}
\left[ {\cal C}^{ij,{\rm dif}}(K)\right]_{\rm loop} &= 
\left< (J^{i,\rmr})^{\rm dif}_2 (K) (J^{j,\rmr})^{\rm dif}_2 (-K) \right> \notag\\
& = (T D_\rmV \tilde{u}_{\sigma})^2  \left<
\le[\lambda_\rmA \partial^i \left(  \lambda_{\rm V} - 2 \i  \psi_\rmV \right)\ri] (K)  \le[\lambda_\rmA \partial^j \left( \lambda_{\rm V} - 2 \i  \psi_\rmV \right) \ri](-K) \right>
\notag\\
& = (T D_\rmV \tilde{u}_{\sigma})^2 \left( \mathcal F^{ij,\rm rr} - 2 \i \mathcal F^{ij,\rm ra} - 2\i \mathcal F^{ij,\rm ar} \right)\,,
\end{align}
where
\begin{align}
\label{eq:calF}
{\cal F}^{ij,\rm rr} &= \left< \lambda_\rmA \partial^i \lambda_\rmV (K) \left( \lambda_\rmA \partial^j \lambda_\rmV \right)(-K) \right> \,,
\\
{\cal F}^{ij,\rm ra} &= \left< \lambda_\rmA \partial^i \lambda_\rmV (K) \left( \lambda_\rmA \partial^j \psi_\rmV \right)(-K) \right> \,,
\quad
{\cal F}^{ij,\rm ar} = \left< \lambda_\rmA \partial^i \psi_\rmV (K) \left( \lambda_\rmA \partial^j \lambda_\rmV \right)(-K) \right> \,.
\end{align}
From eqs.~(\ref{eq:FDT2}) and \eqref{eq:Gra-Gar}, it is straightforward to show that
\begin{align}
{\cal F}^{ij,\rm rr}(K)=\i {\cal F}^{ij,{\rm ra}}(K)+\i {\cal F}^{ij,{\rm ar}}(K)\,,\quad
{\cal F}^{ij,{\rm ra}}(K)=-({\cal F}^{ij,{\rm ar}})^{*}(K)\,. 
\end{align}
We therefore have
\begin{align}
\label{CS-dif}
\le[ {\CC}^{ij,{\rm dif}}(K)\ri]_{\rm loop}=(T D {\tilde u}_{\s})^2\left[-\i {\cal F}^{ij,{\rm ra}}(K)-\i {\cal F}^{ij,{\rm ar}}(K)\right] =2 (T D {\tilde u}_{\s})^2\, {\rm Im}\, {\cal F}^{ij,{\rm ra}}(K)\, . 
\end{align}

On the other hand, the one-loop corrections to the retarded correlator is given by
\begin{align}
\label{CR-final}
\le[ {\cal C}^{ij,{\rm dif}}_{\rm R}(K)\ri] _{\rm loop} &= 
{\rm i} \langle (J^{i,{\rm r}})^{\rm dif}_{2}(K)\, (J^{j,{\rm a}})^{\rm dif}_{2}(-K)\rangle
\no \\
&= \omega T (D_\rmV \tilde{u}_{\sigma})^2 
\langle\le(\l_{\rm A}\pd^{i}(\l_{\rm V}-2{\rm i}\psi_{\rm A})\ri)(K)\, (\l_{\rm A}\pd^{j}\psi_{\rm V})(-K)\rangle
\notag\\
&= \omega T (D_\rmV \tilde{u}_{\sigma})^2 \mathcal F^{ij,{\rm ra}}(K) \,.
\end{align}
Comparing eq.~\eqref{CS-dif} with eq.~\eqref{CR-final}, we immediately verify
\begin{align}
\le( {\CC}^{ij,{\rm dif}}\ri)_{\rm loop}= \frac{2T}{\omega}{\rm Im} \le({\cal C}^{ij,{\rm dif}}_{\rm R}\ri)_{\rm loop}\, . 
\end{align}
Note that if we had ignored the multiplicative noise contributions, i.e., the last two terms in eq.~\eqref{eq:4-sym-cor-dif}, 
we would obtain the wrong relation 
\begin{align}
\left[{\cal C}^{ij, {\rm dif}}(K)\right]_{{\rm loop}}=(T D {\tilde u}_{\s})^2 {\cal F}^{\rm rr}(K)=-\frac{2 T}{\omega}{\rm Im}\left[C^{ij,{\rm dif}}_{\rm R}(K)\right]_{{\rm loop}}\,.
\end{align}
Furthermore, one would also get a wrong relation between $\mathcal C^{00}$ obtained in appendix \ref{sec:FDT-check} and ${\cal C}^{ij}$:
\begin{align}
\lim_{\kv \rightarrow \zev} \frac{1}{2T} \hat {k}^i \hat {k}^j \left[ {\cal C}^{ij,{\rm dif}}(K) \right]_{\rm loop} = - \lim_{\kv \rightarrow \zev} \frac{1}{2T} \frac{\omega^2}{\kv^2} \mathcal C^{00}(K) \,,
\end{align}
which contradicts with eq.~\eqref{eq:Kubo-2} based on the Ward-Takahashi identity.

\bibliographystyle{JHEP.bst}
\bibliography{refs.bib}

\end{document}